\documentclass{ws-jai}
\usepackage{float}
\usepackage{xcolor}
\usepackage{comment}
\usepackage[flushleft]{threeparttable}
\usepackage{subcaption}
\usepackage{natbib}
\usepackage{aas_macros}
\graphicspath{ {02-SysDesc/figs}}
\newcommand{\rfsoci}{\texttt{rfsocinterface}}
\begin{document}

\catchline{}{}{}{}{} 

\markboth{Roberson et al.}{An Open Source RFSoC-based Readout}

\title{An Open Source RFSoC-based Readout Electronics System for Arrays of Kinetic Inductance Detectors and Superconducting Resonators}

\author{Cody Roberson$^{1,*}$, Nia McNichols$^{2}$, Jack Sayers$^{2}$, Sean Bryan$^{3}$, Daniel Cunnane$^{4}$, Peter K. Day$^{4}$, Fabien Defrance$^{4}$, Simon Hempel-Costello$^{2}$, Tracee Jamison-Hooks$^{3}$, Bradley R. Johnson$^{5}$, Maya Lee$^{3}$, Philip Mauskopf$^{3,6}$,  Adrian Sinclair$^{7}$, Liam C. Walters$^{5}$, and Eric Weeks$^{8}$}

\address{
$^{1}$School of Earth and Space Exploration, Arizona State University, Tempe, AZ 85281, USA, carobers@asu.edu\\
$^{2}$California Institute of Technology, Pasadena, CA 91125, USA\\
$^{3}$School of Earth and Space Exploration, Arizona State University, Tempe, AZ 85281, USA\\
$^{4}$NASA Jet Propulsion Laboratory, California Institute of Technology, Pasadena, CA 91011, USA\\
$^{5}$Department of Astronomy, University of Virginia, Charlottesville, VA 22904, USA\\
$^{6}$Department of Physics, Arizona State University, Tempe, AZ 85281, USA\\
$^{7}$Johns Hopkins University, Greenbelt, MD 20771, USA\\ 
$^{8}$Custom Design and Instrumentation, LLC, Tempe, AZ 85284, USA\\
$^{*}$carobers@asu.edu
}

\maketitle

\corres{$^{*}$Corresponding author.}

\begin{history}
\received{(to be inserted by publisher)};
\revised{(to be inserted by publisher)};
\accepted{(to be inserted by publisher)};
\end{history}

\begin{abstract}
We present a novel readout electronics system for large arrays of superconducting electromagnetic resonators, such as kinetic inductance detectors (KIDs), based on a Radio Frequency System on Chip (RFSoC) architecture. Each channel in the readout system is designed for frequency division multiplexing of up to 1024 high quality factor resonances placed on a single microwave transmission line at unique frequencies within a 512~MHz bandwidth. We describe the design of the digital and analog signal processing chains for the implementation of a two-channel 2048-resonator system on the Xilinx ZCU111 RFSoC evaluation board in combination with a custom intermediate frequency (IF) system to convert the RFSoC band to higher frequencies up to 4 GHz. We also detail an associated software interface that provides a range of tools commonly utilized for characterizing KID resonators and for operating them as part of a photometric millimeter-wave imager.  We additionally provide noise characterizations of the individual readout components, along with the complete readout system in isolation and in operation to readout KID resonators. Utilizing millimeter-wave KIDs as a specific example, we find that the performance of the readout system is sufficient for detector-noise-limited operation while also being scalable to the large detector counts in existing and planned imagers. When operated with the maximum number of readout tones, the achieved noise spectrum of the full system is flat down to at least 100~mHz with an amplitude relative to the carrier tone of $-100$~dBc~Hz$^{-1}$. With a smaller number of readout tones, noise amplitudes as low as $-110$~dBc~Hz$^{-1}$ are possible.
\end{abstract}


\section{Introduction}
\noindent Modern astronomical instruments increasingly rely on large-format detector arrays, requiring digital readout electronics capable of simultaneously processing hundreds to thousands of detector channels while maintaining low noise performance. Different detector technologies have motivated several multiplexing architectures. These include sample-up-the-ramp (SUR) readout for HgCdTe detector arrays \citep[e.g., SPHEREx,][]{Heaton2023}, time-division multiplexing (TDM) for transition-edge sensor (TES) arrays (e.g., SPIDER, \citealt{Filippini2010}; BICEP, \citealt{Schillaci2023}), and frequency-division multiplexing (FDM) electronics for TES arrays (e.g., EBEX, \citealt{EBEX2018}; POLARBEAR, \citealt{Hattori2016}). Microwave frequency-division multiplexing has similarly become a prominent readout architecture for kinetic inductance detector (KID) arrays in instruments such as BLAST-TNG \citep{Gordon2016}, Toltec \citep{Wilson2020}, NIKA2 \citep{Bourrion2016}, and the Simons Observatory \citep{Zhu2021, Yu2023}. In addition, FPGA-based electronics are used as backend systems for radio astronomy instruments such as the ALMA correlators \cite{Baudry2012}.

The astronomical community has developed open-source FPGA hardware and firmware to facilitate broader use of these systems and to enable continued development as commercial FPGA technology advances \citep[e.g.,][]{Stefanazzi2022,Smith2024}. Previous KID readout systems utilizing FDM electronics have been implemented utilizing the CASPER Reconfigurable Open Architecture Computing Hardware (ROACH-2) platform together with associated firmware modules \citep{Duan2010,Gordon2016,Sinclair2022,Sinclair2024}. More recently, Xilinx has produced a new generation of Radio Frequency System on Chip (RFSoC) products that integrate FPGA computing fabric with analog-to-digital converters (ADCs) and digital-to-analog converters (DACs), along with an embedded processor offering expanded computational resources and higher data throughput, all within a smaller system size with lower power requirements. Although these developments have significantly advanced KID readout technology, deploying a complete readout system often still requires combining firmware, analog electronics, control software, detector calibration tools, and interfaces that are frequently developed independently and tailored to individual instruments. As detector arrays continue to grow, there remains a need for a unified, modular, and openly available readout platform that integrates these capabilities and scales to the large detector counts required by current and next-generation KID instruments.
 
In this work, we present an RFSoC-based readout electronics system designed for arrays of superconducting resonators, with particular emphasis on KIDs. The system combines FPGA firmware, a custom intermediate-frequency (IF) electronics chain, and a modular Python software framework into a complete end-to-end platform for detector characterization and imaging. The architecture supports the simultaneous readout of up to 2048 resonators across two independent 512 MHz bandwidth channels while providing automated tone generation, detector calibration, telescope control, data acquisition, and downstream processing tools. We characterize the performance of the individual components and of the complete readout system, demonstrating detector-noise-limited operation with millimeter-wave KID arrays developed for the SKIPR instrument \cite{Sayers2025}. Although motivated by SKIPR, the architecture is broadly applicable to a wide range of superconducting resonator instruments requiring scalable, low-noise digital readout electronics.


\section{Performance Requirements}
\label{sec:requirements}

To establish a baseline performance requirement for the readout system, we consider the specific example of SKIPR, a ground-based photometric imager operating at millimeter-wave frequencies \cite{Sayers2025}. A single element of the SKIPR focal plane includes 960 KIDs on a single feedline with resonant frequencies spanning 200--600~MHz. Consequently, the readout system must simultaneously generate and monitor $\simeq 1000$ probe tones distributed arbitrarily across this bandwidth. In addition to supporting a large multiplexing factor, the readout must rapidly retune probe tone frequencies as detector resonances shift in response to variations in optical loading and magnetic field during operation. Given the dither rate of the telescope (1.5$^{\circ}$~s$^{-1}$), along with the full-width at half-maximum of the point spread function (0.089$^{\circ}$), the desired signal band extends to approximately 20~Hz. Nyquist sampling therefore requires a readout rate of $\gtrsim 40$ samples s$^{-1}$.

The cryogenic readout circuit in SKIPR has been designed so that, in addition to irreducible noise due to random photon arrivals from the background and intrinsic detector noise, the noise performance is limited by the SiGe low noise amplifier (LNA) located immediately after the detector array on the 3~K cryogenic stage. The noise temperature of the LNA is approximately 2~K, which corresponds to $\simeq -195~dBm ~ Hz$$^{-1}$. For the typical SKIPR detector, the maximum readout power is approximately $-90$~dBm, and in operation the characteristic resonance depth is approximately 5~dB. Thus, referenced to the typical carrier amplitude of the tone, the noise contributed by the SiGe LNA is $\simeq -100$~dBc~Hz$^{-1}$. For context, noise due to the random arrival of photons in typical observing conditions, when referenced to the carrier amplitude, corresponds to $-90$~dBc~Hz$^{-1}$. To ensure that the readout system negligibly impacts the noise performance of SKIPR, we thus require a per-tone noise level of $\lesssim -100$~dBc~Hz$^{-1}$ when operating with 1000 tones. While these requirements are derived from SKIPR, they are comparable to those of other KID-based millimeter-wave and submillimeter imagers \citep[e.g.,][]{Bourrion2016,Adam2018,Paiella2019,Rowe2023,Reyes2026}, and so they provide a reasonable baseline for a wide range of applications.

\section{System Architecture}
\begin{figure}
    \centering
    \includegraphics[width=0.6\linewidth]{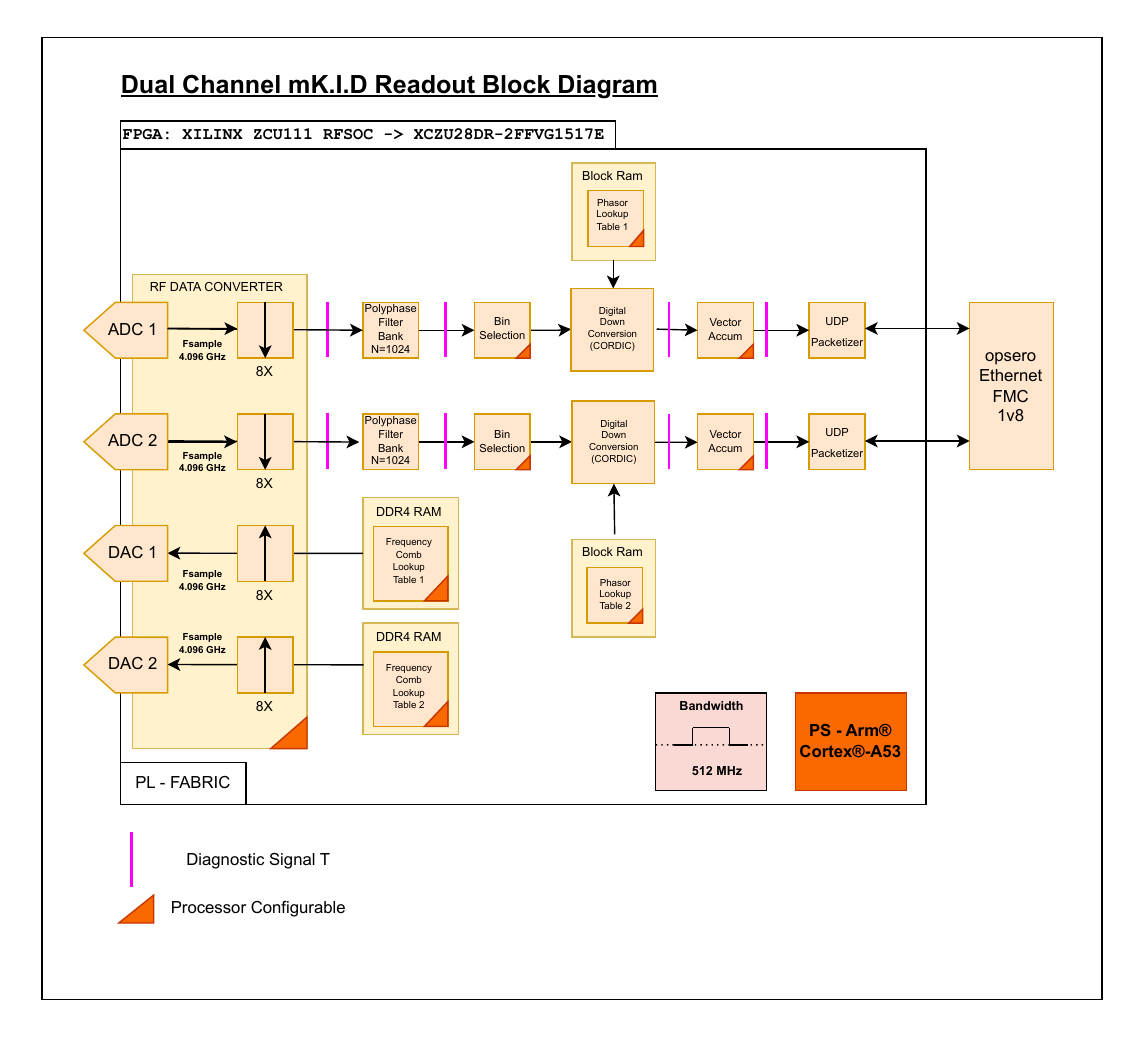}
    \caption{Schematic of the high level design of the firmware.}
    \label{fig:overall_diagram}
\end{figure}

The core of the 
readout system is 
the Xilinx/AMD 
RFSoC architecture, 
which integrates a Processing System (PS) and Programmable Logic (PL, see Figure~\ref{fig:overall_diagram}).
Both the PS and PL are equipped with 4 GB of dedicated, independent DDR4 memory resources. The PS hosts a Yocto built Ubuntu Linux distribution on 
the ZCU111 evaluation board's quad-core ARM Cortex-A53 processor. System level interfacing and
hardware software interaction are facilitated by PYNQ, an open-source framework by AMD that
provides Python based libraries for 
interaction with the programmable logic. 
Hardware designs compiled within AMD Vivado that utilize predesigned Intellectual Property (IP) 
blocks are supported by the PYNQ framework. When high level PYNQ functions are 
executed, data is transferred between the PS and PL via the Advanced eXtensible Interface (AXI) 
protocol. Upon loading an FPGA firmware, referred to as an 'overlay', the PYNQ 
framework dynamically parses the hardware metadata to expose software interfaces and 
drivers. This abstraction layer significantly accelerates development cycles, 
and serves as the software backbone for our 
readout system.
A free open-source library called \texttt{KidPy} was developed to facilitate hardware configuration 
and control for the end user. Redis, a service offering both key-value stores as well
as publish / subscription messaging provides the commanding infrastructure to the instrument. 
Within the PS of the RFSoC's, a redis client polls for messages published from the host computer and
processes the request, interacting with the hardware as needed. Finally, the client replies to the host
with the results of the request. The following sections describe the major components of the readout architecture, including tone generation, detector readout, data collection, the IF electronics, and the host software framework.

\subsection{Tone Generation}
\texttt{Kidpy} provides high-level Application Programming Interface (API) procedures to initialize 
and synthesize the target probe tone waveforms. Previous generations of this readout system, developed for BLAST-TNG by 
\citet{Gordon2016}, performed baseband Direct Digital Synthesis (DDS) of the 
readout probe tones directly within the FPGA's integrated Block RAM (BRAM). While effective, the underlying programmable fabric 
provided a total of 38~Mb (approximately 4.75~MB) of available BRAM, fundamentally limiting the maximum number 
of simultaneous probe tones that could be generated. This architecture imposed a strict tradeoff between total 
bandwidth and frequency resolution, requiring tone placement to be heavily prioritized around the detectors'
resonant frequencies. To overcome these limitations, subsequent developments for CCAT-Prime-Cam \cite{Sinclair2022} 
utilized external DDR4 memory rather than internal BRAM for waveform generation. The current readout adopts this architecture, significantly alleviating BRAM resource constraints and enabling the simultaneous 
readout of two detector tiles per RFSoC.

To generate the waveforms, a frequency-domain lookup table (LUT) consisting of $2^{20}$ Fast Fourier Transform (FFT) 
bins is allocated. For each tone in the desired set of probe tones, a pseudo-random phase drawn from a uniform distribution 
is generated and assigned to the nearest corresponding FFT bin. The frequency-domain LUT is subsequently transformed into a 
time-domain data stream via an Inverse FFT (IFFT). The resulting data are scaled to match the 14-bit resolution 
of the 
digital-to-analog converter (DAC). Finally, the waveform data are written to the PL DDR4 memory using 
the PYNQ \texttt{MMIO} library. Within the firmware domain, these values are continuously retrieved from the DDR4 memory
and routed to the AMD RF Data Converter (RFDC) block. The RFDC interpolates the data by a factor of 8, 
generating two pairs of analog in-phase and quadrature (I/Q) signals at a sampling rate of 4.096~GHz. 
These final analog signals are then routed to an external, 1U-sized intermediate frequency (IF) up/down converter, 
where they are mixed with a carrier tone to 
modulate them into to the frequency band of the resonator detectors.

\subsection{Detector Sensing}
\begin{figure}
    \centering
    \includegraphics[width=0.6\linewidth]{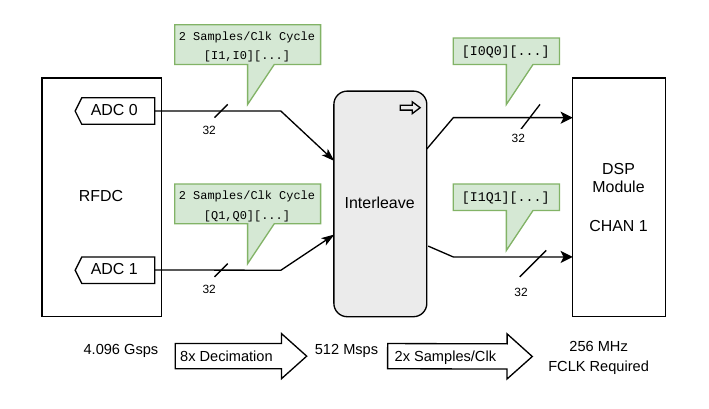}
    \caption{ADC AXIS datapath}
    \label{fig:adcDataPath}
\end{figure}

After probing the resonator detectors, the analog signals return to the IF electronics, where they are down-converted into I/Q signals before being sampled by the RFSoC's analog-to-digital converters (ADCs). The ADCs operate at a sampling rate of 4.096~GSPS. To reduce the data rate and satisfy both timing and processing constraints, the signals are decimated by a factor of 8 within the RFDC block. The resulting digitized data stream consists of two I/Q signals, each containing two samples of data per clock cycle (Figure \ref{fig:adcDataPath}).

The core of the detector readout is a custom Digital Signal Processing (DSP) IP core developed within the 
\texttt{mlib\_devel} environment, an open-source hardware design platform maintained by the Collaboration 
for Astronomy Signal Processing and Electronics Research (CASPER). Input data streams are initially routed into 
a 1024-point ($2^{10}$), four-tap Polyphase Filter Bank (PFB) employing a Finite Impulse Response (FIR) structure 
with a Blackman-Harris windowing function. The filtered output is subsequently passed to a FFT core that preserves the two-sample-per-cycle architecture required for real-time throughput. 
The output is subsequently streamed into a FFT core that similarly processes two 
parallel samples per cycle to maintain real-time throughput. Following the FFT, the data stream is routed to a 
channel-selection and a double-buffered (ping-pong) memory module. During the tone generation stage, the \texttt{kidpy} 
control software calculates which firmware FFT bins correspond to valid probe tone frequencies. 
The channel-selection module then dynamically selects these specific FFT bins from that list while discarding unutilized 
channels.

In legacy firmware iterations, two 32-bit BRAM blocks with a depth of $2^{18}$ were utilized to 
implement a secondary LUT containing 1024 tones, each containing 512 samples. 
This LUT data was fed concurrently with the reordered FFT data into a digital down-conversion (DDC) stage, 
shifting the signals from a 500~kHz intermediate frequency to a baseband rate of approximately 1~kHz. 
Following the approach demonstrated in \cite{Sinclair2022} for the current dual-channel firmware, 
the BRAM-based LUT was replaced by two AMD CORDIC IP blocks to perform the down-conversion. 
This modification yields a substantial reduction in BRAM utilization, at the cost of rendering the firmware 
incompatible with TES detector readouts. Following the DDC stage, the channelized data 
streams are accumulated and written to a buffer. A dedicated Ethernet module 
supporting the User Datagram Protocol (UDP) packages the data. The module prepends a header containing user 
parameters configured via \texttt{kidpy}, reads the accumulated data from the buffer, and transmits the 
1024 interleaved I/Q channels via Ethernet to the primary host computer controlling the full system.
Included in this dataset is a packet count, and a binary flag that indicates whether or not a Pulse Per Second (PPS)
signal was detected at the RFSoC. This provides a way to synchronize data across multiple external subsystems. 

\subsection{Data Collection}
The final component of the readout architecture is the data collection framework, which records detector data together with the metadata required for downstream processing. The readout software 
utilizes the HDF Group's Hierarchical Data Format 5 (HDF5) library to manage both pre- and post-processed 
detector data. During data acquisition, \texttt{kidpy} employs multiprocessing to simultaneously record data 
streams from one or more RF chains, with each RF chain allocated a dedicated HDF5 file. Readout configurations 
and collection metadata are stored within a `Global Data' group, whereas the core detector data, timestamps, 
and PPS events are organized into a `Time Ordered Data' group.

\subsection{IF electronics}
\begin{figure}[ht!]
    \centering
    \includegraphics[width=0.6\linewidth]{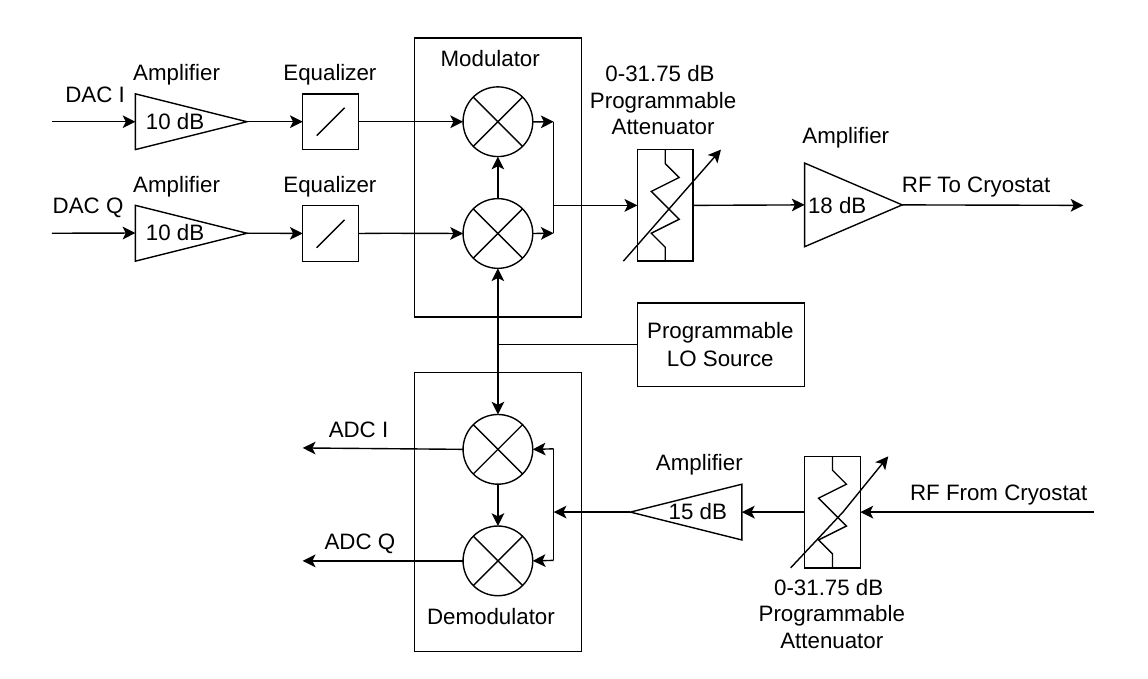}
    \caption{Schematic of the main components of the IF converter.}
    \label{fig:IF}
\end{figure}

With the architecture detailed above, the DACs are able to output tones at frequencies up to 256~MHz. To facilitate measurement of resonators at higher frequencies, along with enabling rapid frequency sweeps of the resonators for calibration, additional IF electronics were developed as part of the overall readout system. As shown in Figure~\ref{fig:IF}, the IF electronics include analog quadrature modulators to up-convert the DAC-generated tones prior to probing the resonators and to down-convert the tones prior to digitizing the signal with the ADCs. A single programmable LO source, capable of generating a reference at frequencies up to 4~GHz, provides the modulation for both the up and down conversion. By utilizing a homodyne mixing technique, phase noise in the LO is largely canceled, significantly improving the end-to-end performance of the readout system (see Section~\ref{sec:IF_noise}).

\subsection{Rfsocinterface}

\begin{figure}
    \centering
    \begin{subfigure}[c]{0.48\textwidth}
        \includegraphics[width=\linewidth]{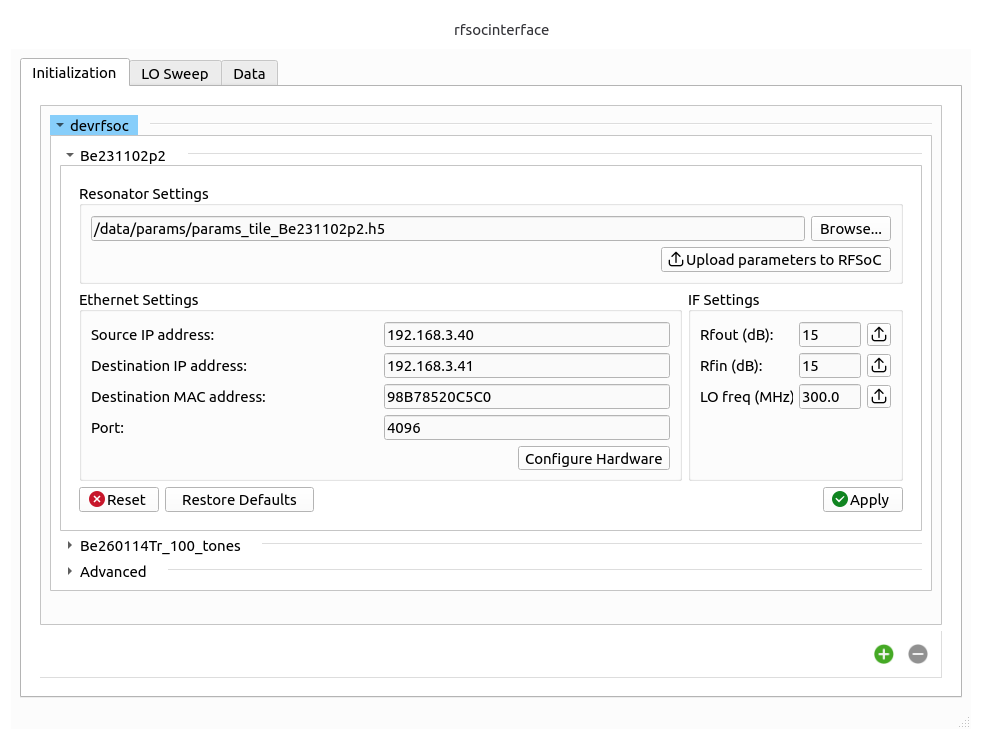}
        \caption{Initialization Tab}
        \label{fig:gui-init}
    \end{subfigure}
    \begin{subfigure}[c]{0.5\textwidth}
        \includegraphics[width=\linewidth]{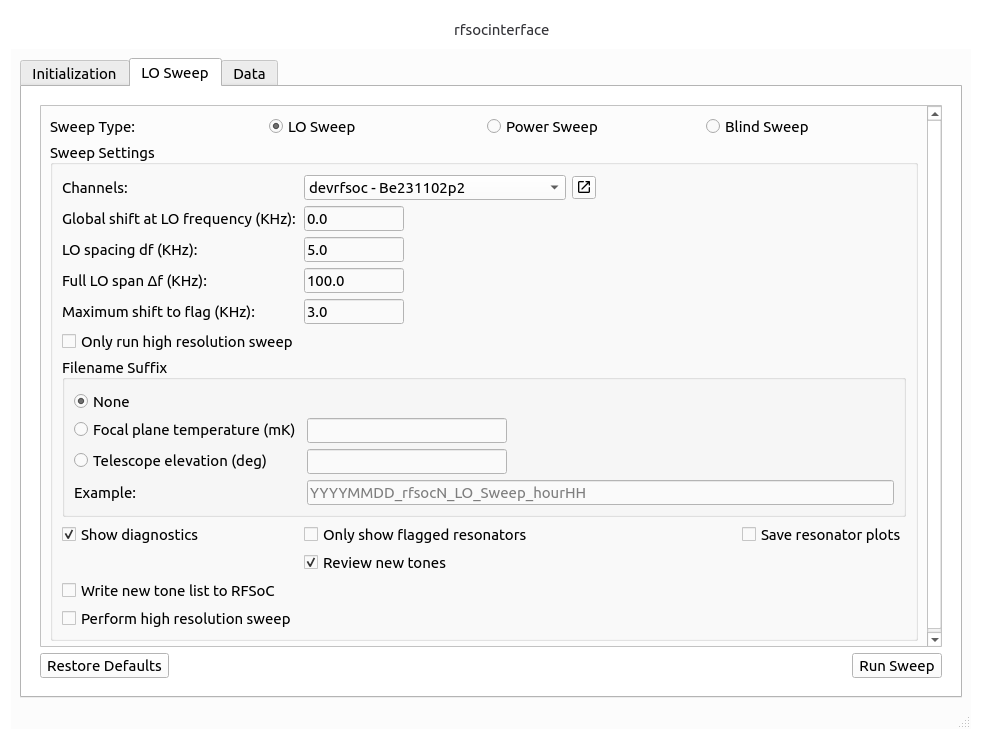}
        \caption{Calibration Tab}
        \label{fig:gui-calibration}
    \end{subfigure}
    \begin{subfigure}[b]{0.48\textwidth}
        \includegraphics[width=\linewidth]{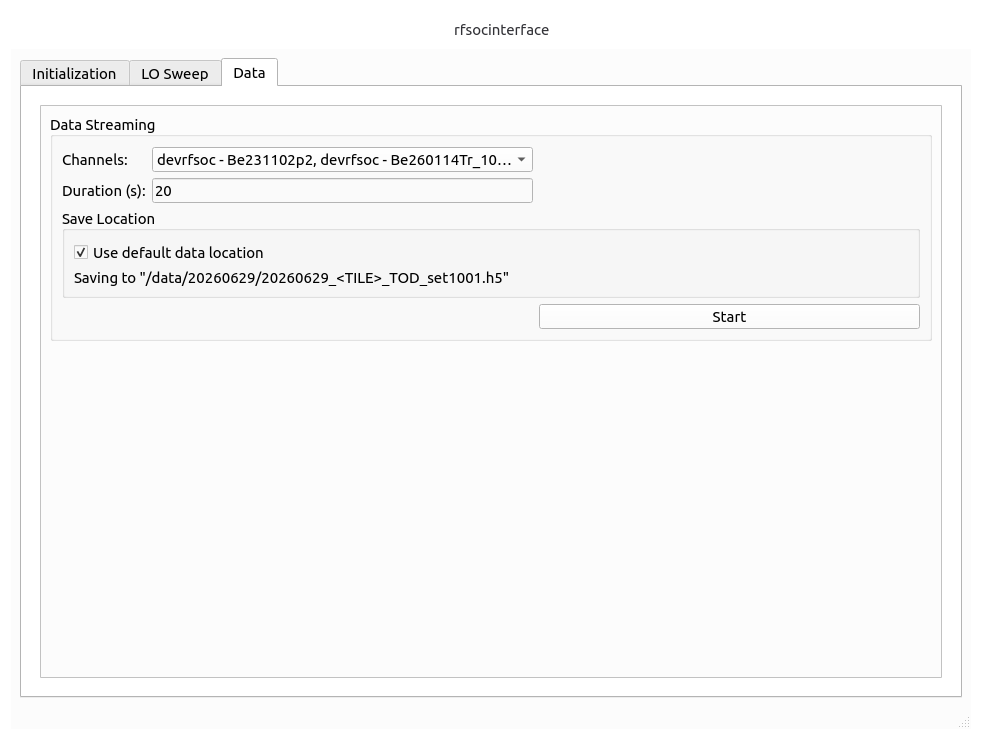}
        \caption{Data Collection Tab}
        \label{fig:gui-data}
    \end{subfigure}
    \begin{subfigure}[b]{0.5\textwidth}
        \includegraphics[width=\linewidth]{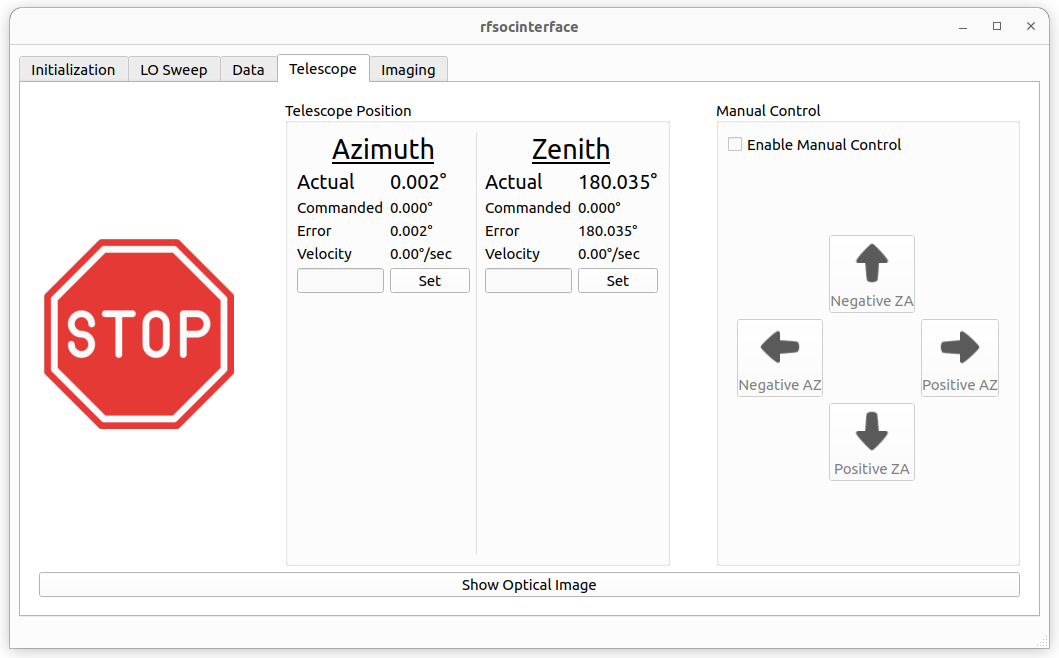}
        \caption{Telescope Control Tab}
        \label{fig:gui-tele}
    \end{subfigure}
    \caption{Tabs from the \rfsoci\ GUI}
    \label{fig:gui}
\end{figure}

We developed \rfsoci, a Python software library, to handle the collection and processing of data from the RFSoC readout. \rfsoci\ was designed with flexibility and ease-of-use at the forefront. It is highly modular,  allowing users to easily adapt the software to their specific needs without exiting the program. While \rfsoci\ can be imported into other projects like any typical Python library, it also includes a graphical user interface (GUI), that provides a streamlined way to interface with the RFSoC. From the GUI, users can easily reconfigure the RFSoC as needed, collect data, reposition a telescope, and adjust downstream data processing. These features are grouped into different tabs to provide an intuitive workflow, see Figure~\ref{fig:gui}. Below, we describe each of these tabs and their associated functions.

\subsubsection{Initialization} The initialization tab serves as a hub for reconfiguring the RFSoC system, see Figure~\ref{fig:gui-init}. Here, the user can select a new list of baseband tones for playback, adjust the LO frequency for up/down-conversion, and change the power levels of the variable attenuators in the IF system. While the aforementioned features are used frequently, there are additional configuration parameters that rarely change. These include the IP address of the RFSoC, the Ethernet and USB ports used to connect to the readout, and the firmware bitstream used for the RFSoC FPGA. As these settings are rarely changed, they are hidden in the ``advanced'' sub-tab to avoid cluttering the screen while still readily providing the essential functionality.

\subsubsection{Calibration} 
This tab deals with the calibration of the RFSoC probe tones, see Figure~\ref{fig:gui-calibration}. KIDs are sensitive to environmental changes, and their resonant frequencies can shift from their nominal positions. To account for this, we perform an LO sweep to identify these shifts and to correct our list of tones accordingly, thus calibrating the readout for the local environment at the time of data capture. 

Before the LO sweep begins, the user selects a tone list that serves as an estimate of the actual resonance locations. In general, this is the most recently calibrated tone list, but it could also be a default tone list for a given detector array. Next, the LO frequency in the IF system is stepped over a user-specified range with a user-specified step size. We record the complex-valued transmission of each tone at each LO frequency. While these data could be fitted to a superconducting resonance model for each KID, for computational speed the default operation instead relies on fitting a quadratic function to the transmission magnitude. The frequency corresponding to the minimum of this quadratic fit is then assumed to be the resonant frequency, and a candidate tone list based on these positions is generated.

\begin{figure}
    \centering
    \includegraphics[width=\linewidth]{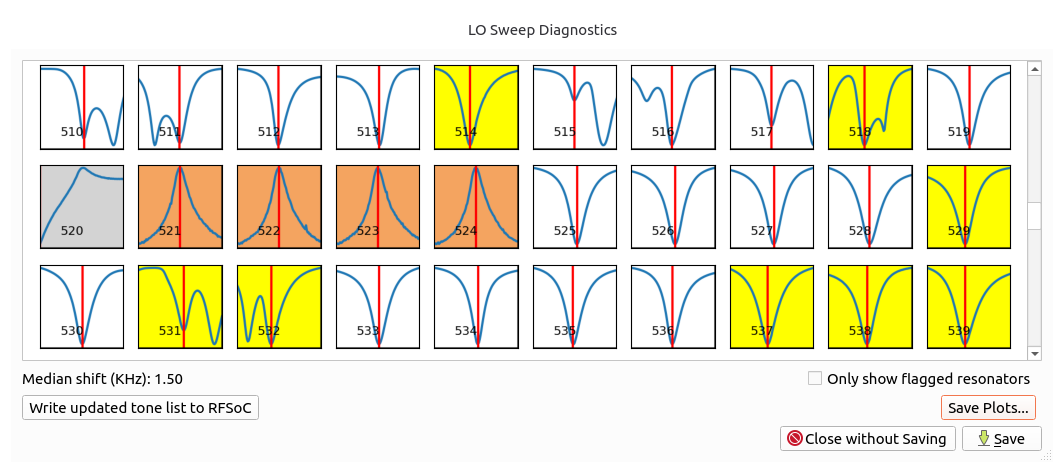}
    \caption{Example LO sweep plots. For each tone, the transmission magnitude is plotted in blue with a vertical red line indicating the location of the fitted resonance frequency. Tones that have shifted far from their original position are flagged with a yellow background. Off-resonance tones are indicated by an orange background. Tones that have been denoted as non-opertaional are indicated by a gray background.}
    \label{fig:LO}
\end{figure}

The simple quadratic fit, while fast, can sometimes fail to locate the correct resonance frequency, or, in the case of nearby resonances, assign multiple tones to the same frequency. Both scenarios require user input to amend. To facilitate this, the GUI displays a plot of the transmission magnitude with a vertical line indicating the location of the fitted resonance frequency for each tone, see Figure~\ref{fig:LO}. The GUI also flags tones that have shifted from their original position in excess of a user-defined threshold. It highlights the plots associated with these tones, allowing the user to quickly examine the LO sweep results and focus on problematic tones. The user can interact with this plot to adjust incorrect tones. This involves dragging the vertical line to the desired resonance frequency, with the option to either set the tone frequency directly to this new value or reinitialize the quadratic fit from that starting location. In the latter case, the location of the updated tone frequency is displayed, allowing the user to confirm it is in the correct location, and to iterate if needed.

In addition to the LO sweep, this tab provides functionality for other tone calibration procedures. When the initial tone list is unknown, we perform an uninformed sweep. In this case, the LO frequency is again stepped over a user-specified range with a user-specified spacing, but using a list of equally spaced tones distributed across the entire bandwidth of the readout. By default, the range of this sweep is set to be larger than the tone spacing to provide overlapping data. As with the targeted LO sweep detailed above, the complex-valued transmission of each tone is converted to a transmission magnitude, and negative excursions relative to the baseline are identified. The parameters related to this identification, for instance, an excursion threshold , are specified by the user. The GUI again provides an interactive plot to examine the full set of identified resonance frequencies and adjust as needed, noting that multiple (or no) resonances may fall within the sweep of any given tone.

The last type of calibration sweep is a power sweep, used to determine the optimal power for each probe tone. In a power sweep, we step the overall power by adjusting the variable attenuators in the IF system and conduct an LO sweep centered on each resonator at each power level. The same quadratic fits utilized in the targeted LO sweeps are used to determine the resonance frequency of each KID at each power level. At low power, the resonant frequency is independent of readout power. At higher readout powers, the resonance frequency decreases with increasing power. In general, it is desirable to set the readout power as high as possible while minimally impacting the resonance, and the user can specify a maximum frequency shift relative to the low-power baseline to obtain this value for each KID from the automated analysis.

\subsubsection{Data Collection and Processing}

The \rfsoci\ GUI contains two tabs for data collection, named data (see Figure ~\ref{fig:gui-data}) and imaging. The data tab allows the user to capture RFSoC data for a specified period of time. Similarly, the imaging tab allows the user to collect data, but with the express purpose of generating an image from SKIPR. The user is able to select from a set of predefined telescope dithering patterns to execute while data is being streamed. In the future, the user will also be able to write their own dither patterns and import them into the GUI from this tab.

Data are automatically stored in HDF5 files on the local file system. In addition to the raw complex-valued transmission data recorded from the RFSoC for each tone (referred to as I/Q basis), these files contain data transformed to the gain/phase basis, frequency/dissipation basis, and calibrated to brightness units (mK). In the case of imaging, the files also contain telescope pointing information during the observation. Additionally, these data files contain several important parameters needed for downstream processing. These include but are not limited to: the selected tone list at the time of data capture, the most recent LO sweep, detector positions relative to the telescope pointing center, polarization selectivity, and a list indicating the status of each probe tone. This status list indicates whether a given tone is centered on a KID resonance or placed at an off-resonance position and, for the former, whether the KID is denoted as operational or non-operational. KIDs may be designated as non-operational due to significant cross-talk with nearby detectors, high noise, low optical response, or any other reason. These data files are designed to allow the user to easily perform any necessary processing.

\rfsoci\ contains several tools for processing data. Processing is divided into different routines that affect the data in various ways. All of these routines are designed to be applied in any order and readily rearranged in the GUI, providing flexibility in how data is processed. Some of these routines are applied to the time-ordered detector data, including both high-pass and low-pass filters, digital down-sampling, flagging of cosmic-ray events, and correlated noise removal. Other routines are applied to the reconstructed images, such as Gaussian blurring to suppress noise on small angular scales. Additionally, the library provides a simple API for developing new routines, allowing users to easily add new processing steps as needed.

While streaming data from the RFSoC, data packets are timestamped upon their arrival at the host computer. Despite collecting data at a regular sampling rate of 488 Hz within the RFSoC, there is variability in how long each sample takes to arrive at the host computer, resulting in jitter in timestamp values. Additionally, the host computer is connected to the RFSoC via UDP to maximize throughput. However, UDP does not guarantee that each packet will reach the processing computer, potentially creating gaps in the data stream. In practice, approximately 1 in $10^4$ packets fail to reach the host computer. However, to facilitate downstream processing, for instance, the application of a high-pass filter, we require equally spaced, continuous time-series data. To address this, the RFSoC records a count index and transmits it to the host computer with each data packet; any gap in this index, therefore, corresponds to a lost packet. After identifying any lost packets via this index, we then fit a degree-four polynomial to the nearby samples to fill in values for the missing detector data, followed by a linear interpolation to create corrected timestamps with uniform spacing. We note that this latter interpolation affects only the timestamp data, not the detector data.

When constructing images, we must ensure that the telescope pointing information is properly synchronized with the detector data recorded by the RFSoC. The RFSoC and the telescope motor position encoders have different sample rates, with the latter operating at approximately 100~Hz, and so the telescope position data is up-sampled to match the RFSoC data. Because the telescope motion is smooth, this up-sampling accurately captures the position at the higher data cadence of the RFSoC. After this up-sampling, we empirically determined that there is an offset between the timestamps of the RFSoC data and the corresponding packets from the telescope motor controller based on rapid dithers in opposite directions past a bright calibration source. The offset is not constant and is typically on the order of 10 samples, or approximately 20~ms. To reconcile this, a single one pulse per second (PPS) source is connected to the RFSoC and both the azimuth and zenith angle motor encoders to provide a common absolute time reference. Because the typical offset is much shorter than the spacing between pulses, it is straightforward to align the detector and telescope datastreams using the digitized pulse arrivals.

\subsubsection{Telescope Control}

The telescope control tab allows the user to reposition SKIPR from the GUI, see Figure~\ref{fig:gui-tele}. The primary mode of repositioning is to input the desired azimuth and zenith angle of the telescope, after which \rfsoci\ will automatically shift the telescope to the desired location. Should the user instead want to move the telescope in a more interactive manner, the GUI also provides a directional pad to manually adjust the angles in real-time. At all times, the current position, the desired position, the difference between them, and the rate of travel are displayed in the GUI. In addition, the live stream from an optical-wavelength camera co-aligned with the telescope bore-sight is displayed, providing a quick visual reference of the current pointing.

\section{System Noise Performance}

\subsection{RFSoC DAC and ADC characterization}
\label{sec:DAC_noise}

\begin{figure}
    \centering
    \includegraphics[width=0.48\textwidth]{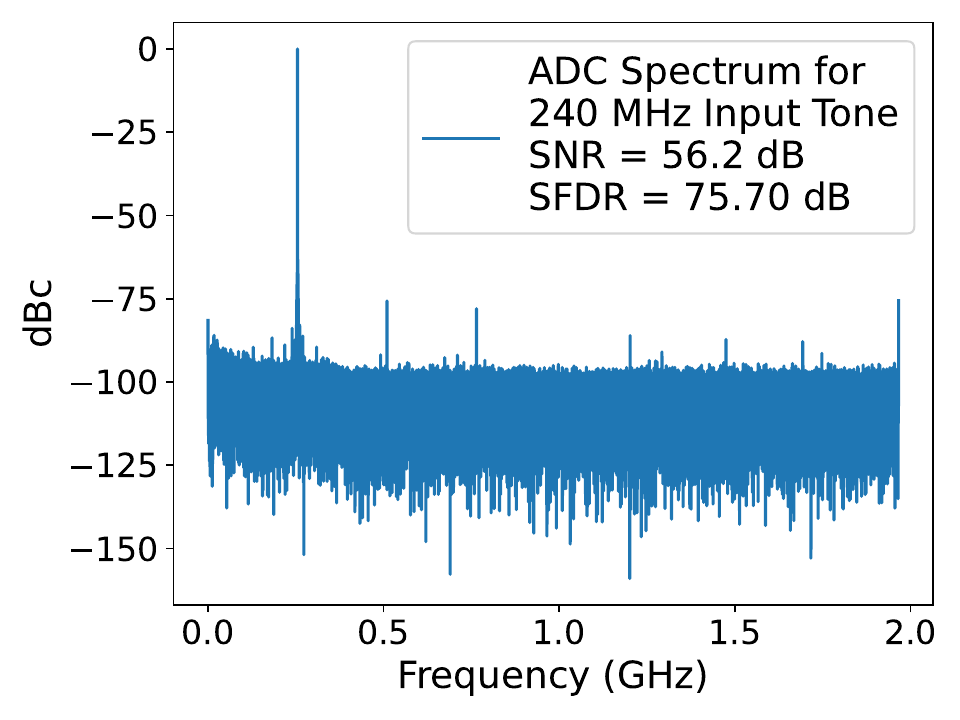}
    \includegraphics[width=0.50\textwidth]{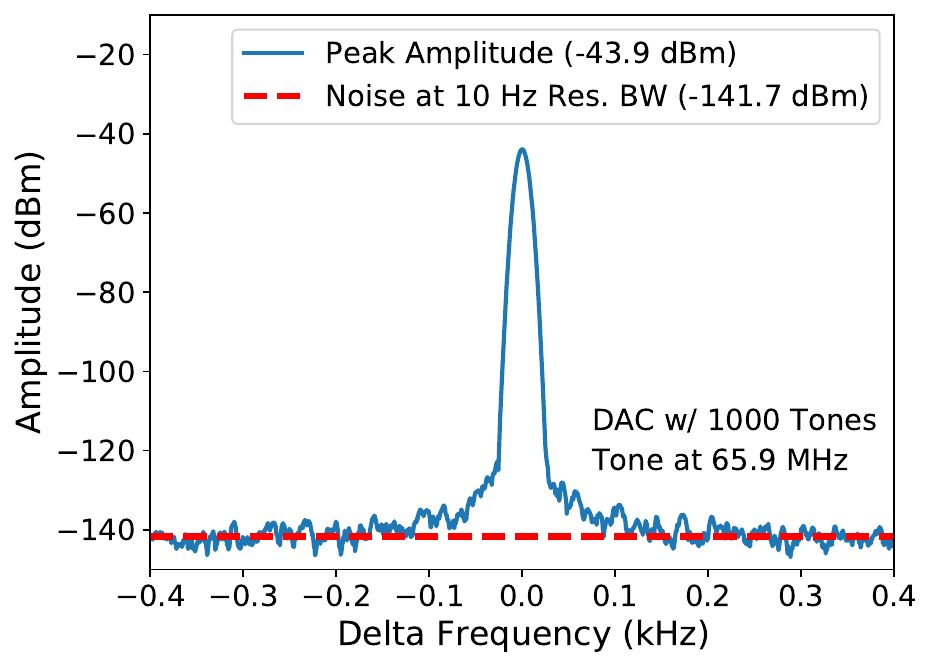}
    \caption{Left: measured performance of the ADCs, obtained by digitizing a single tone centered at 240~MHz with an amplitude of $+5$~dBm generated by an external synthesizer. At a sample rate of approximately 4~GHz the SNR is equal to 56.2~dB, implying digitization noise of $-149.2$~dBm~Hz$^{-1}$. Right: measured performance of the DACs, obtained by generating 1000 evenly spaced tones and recording one of those tones with an external spectrum analyzer. At 10~Hz resolution bandwidth, the noise is $-141.7$~dBm, implying a digitization noise of $-151.7$~dBm~Hz$^{-1}$.}
    \label{fig:ADC_SNR_SFDR}
\end{figure}

To measure the performance of the RFSoC's ADCs, we developed custom firmware to extract the raw ADC data from the FPGA. The ADC was configured for real (not complex) sampling with the mixer in bypass mode, a decimation factor of $\times$1, and an output rate of 3.932~GHz. We chose this sample rate, rather than the 4.096 GHz sample rate the DACs are capable of, to match the value utilized in the Xilinx white paper on converter performance in order to provide a direct comparison. To perform the measurement, a Rhode \& Scharz SMB100A microwave signal generator was connected to the ADC input. The generator was configured to output a 240 MHz tone at an amplitude of $+5$ dBm, which was empirically determined to be the maximum signal amplitude before clipping. From this configuration, the digitized SNR was measured at 56.2~dB, see Figure~\ref{fig:ADC_SNR_SFDR}. Given the sample rate, this corresponds to a digitization noise of $-149.2$~dBm~Hz$^{-1}$. We additionally note that the spur-free dynamic range was found to be 75.7~dB. All of these values are in good agreement with the published specifications from Xilinx, and suggest the ADCs are performing as expected. Referenced to the performance requirements listed in Section~\ref{sec:requirements}, the maximum per-tone amplitude that could be input to the ADCs when operating with 1000 tones is approximately $-25$~dBm. Given the measured digitization noise, this suggests that noise levels as low as $-124$~dBc~Hz$^{-1}$ are possible, and are well below the requirement of $\lesssim -100$~dBc~Hz$^{-1}.$

The RFSoC's DACs were tested using the standard tone-generation algorithm described above to generate 1000 evenly spaced, approximately equally powered tones. This output was then measured with a Rohde \& Schwarz FSP spectrum analyzer. To obtain a power level just below the spectrum analyzer's saturation, an amplifier with $+12$~dB of gain and a noise figure of $+4$~dB was included in the circuit. Subtracting the gain of this amplifier to obtain values referenced to the DAC output, the typical single-tone power was approximately $-44$~dBm and the off-tone noise was $-151.7$~dBm~Hz$^{-1}$, see Figure~\ref{fig:ADC_SNR_SFDR}. This indicates that, relative to the typical tone carrier power of $-44$~dBm, single-tone noise performance near $-108$~dBc~Hz$^{-1}$ is thus possible from the combination of the DACs when outputting 1000 total tones, approximately 8~dB better than the performance requirement listed in Section~\ref{sec:requirements}.  

\subsection{Digital Loopback}

By connecting the output of the RFSoC DACs directly to the RFSoC ADCs, we are able to assess the noise due solely to the digital electronics. We refer to this operational state as digital loopback. Our baseline operation in digital loopback consists of a total of 1000 tones with equal power and approximately equal spacing within the 512~MHz bandwidth, excluding the edges of the band. Specifically, 500 tones are uniformly spaced between $-246$~MHz and $-11$~MHz, and 500 tones are uniformly spaced between $+10$~MHz and $+245$~MHz. It should be noted that other configurations were also tested, such as 1000 tones placed between $+5$~MHz and $+205$~MHz, and the noise results from these other configurations are consistent with the baseline results described below. 

\begin{figure}
    \centering
    \includegraphics[width=0.49\textwidth]{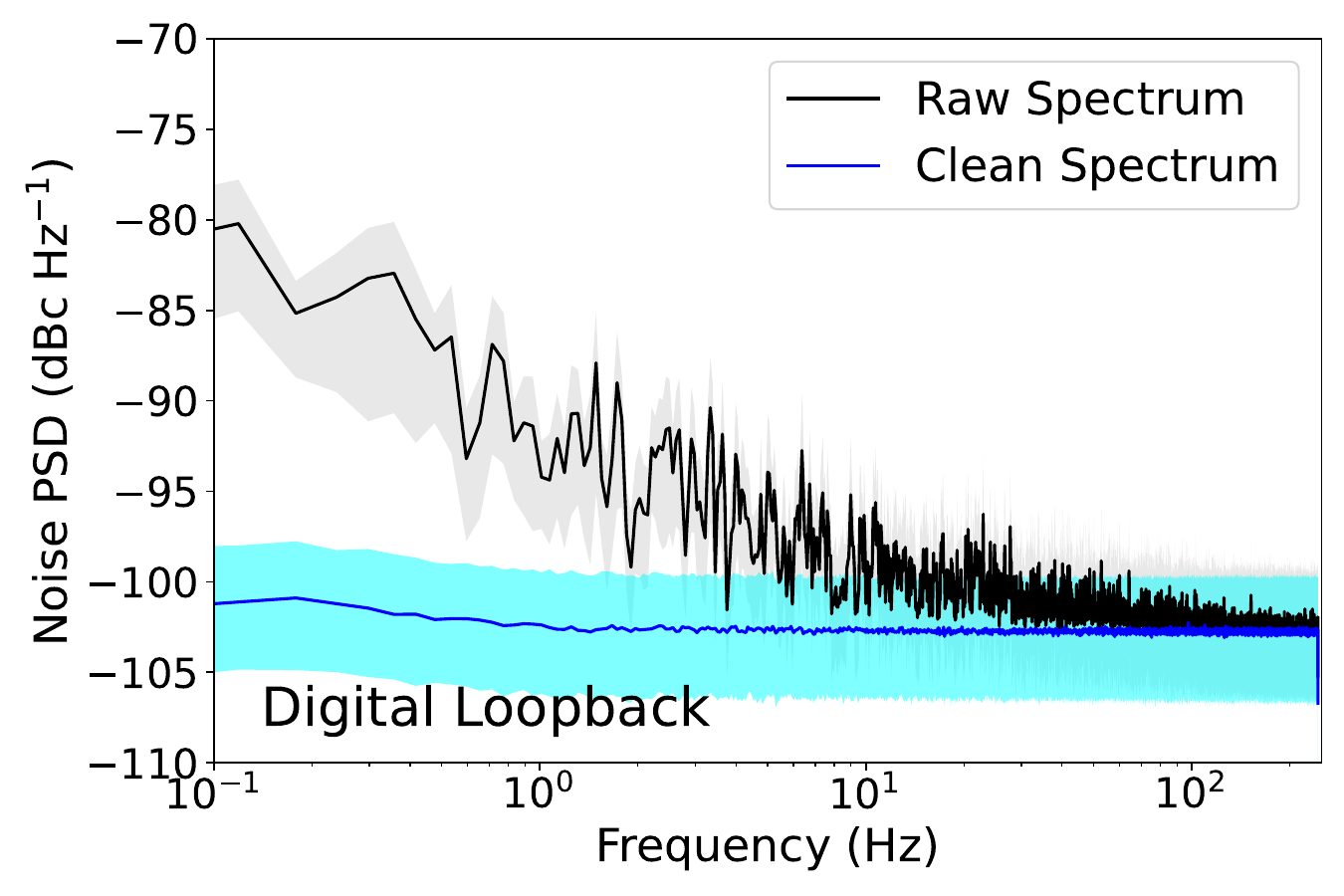}
    \includegraphics[width=0.49\textwidth]{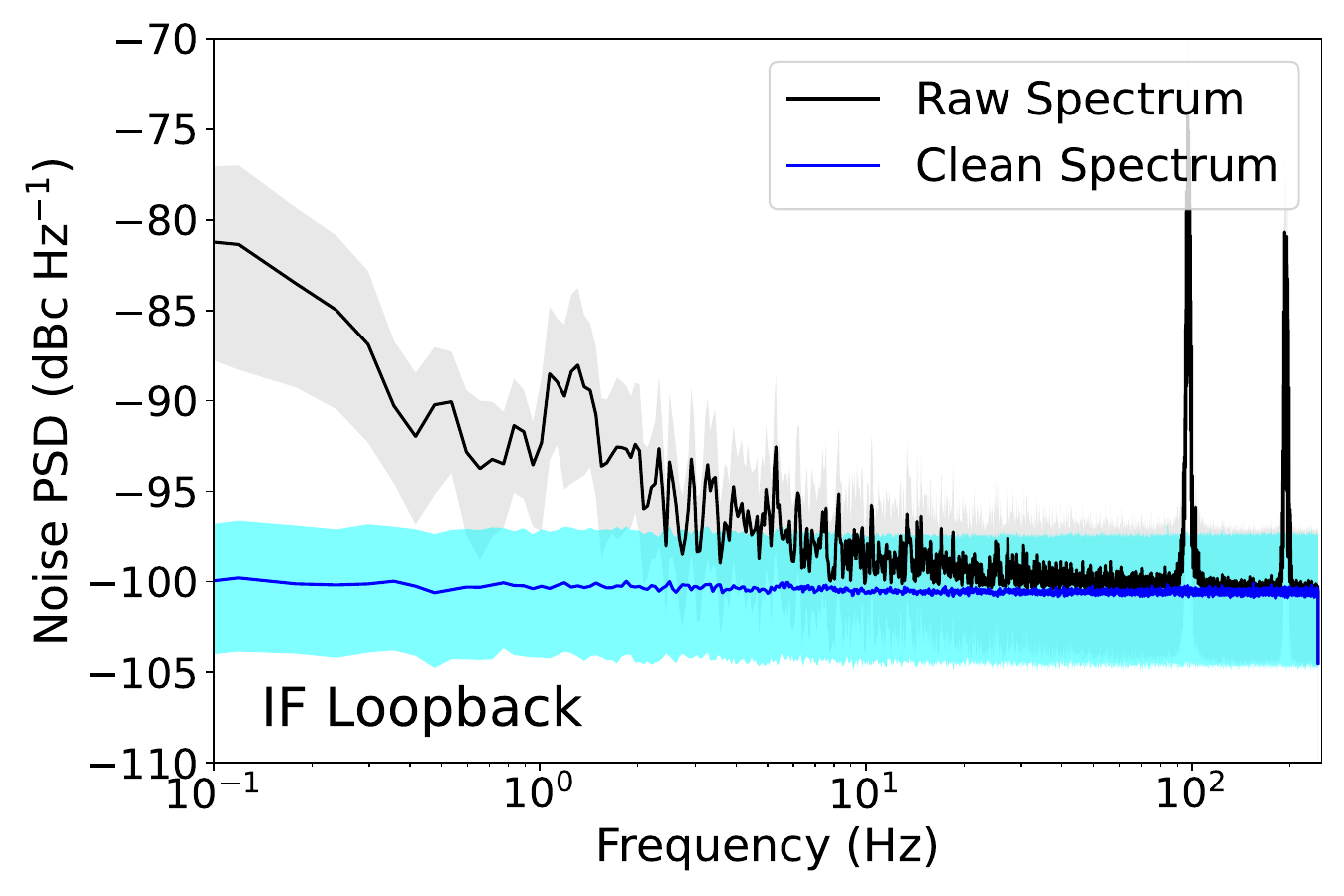}
    \caption{Noise performance of the readout system in digital loopback (left) and IF loopback (right) with 1000 tones. Solid lines denote the median tone performance, and shaded areas enclose 68 percent of the tones. Black corresponds to the raw data collected by the readout system, and blue corresponds to the same data after removing correlated noise.}
    \label{fig:raw_psd}
\end{figure}

A total of 100~seconds of data is collected, and the noise power spectral density (PSD) is computed for each tone based on the average from independent 10-second-long blocks extracted from the full dataset. We then determine the median noise PSD, along with the range spanning 68 percent of the tones, see Figure~\ref{fig:raw_psd}. The raw data show a rising noise PSD at low frequency, increasing from approximately $-100$~dBc~Hz$^{-1}$ at 100~Hz to approximately $-80$~dBc~Hz$^{-1}$ at 0.1~Hz. 

However, this low-frequency noise is highly correlated between tones, and it can thus be readily subtracted. To do so, we construct two templates spanning the full 100~seconds of data, obtained from the two highest-amplitude components of a principal component analysis \cite{Jolliffe1986}. For each tone, we determine the two correlation coefficients with each of these templates, which are then utilized to subtract them from the raw data. The resulting noise PSD is approximately flat over the full bandwidth.

\begin{figure}
    \centering
    \includegraphics[width=0.49\textwidth]{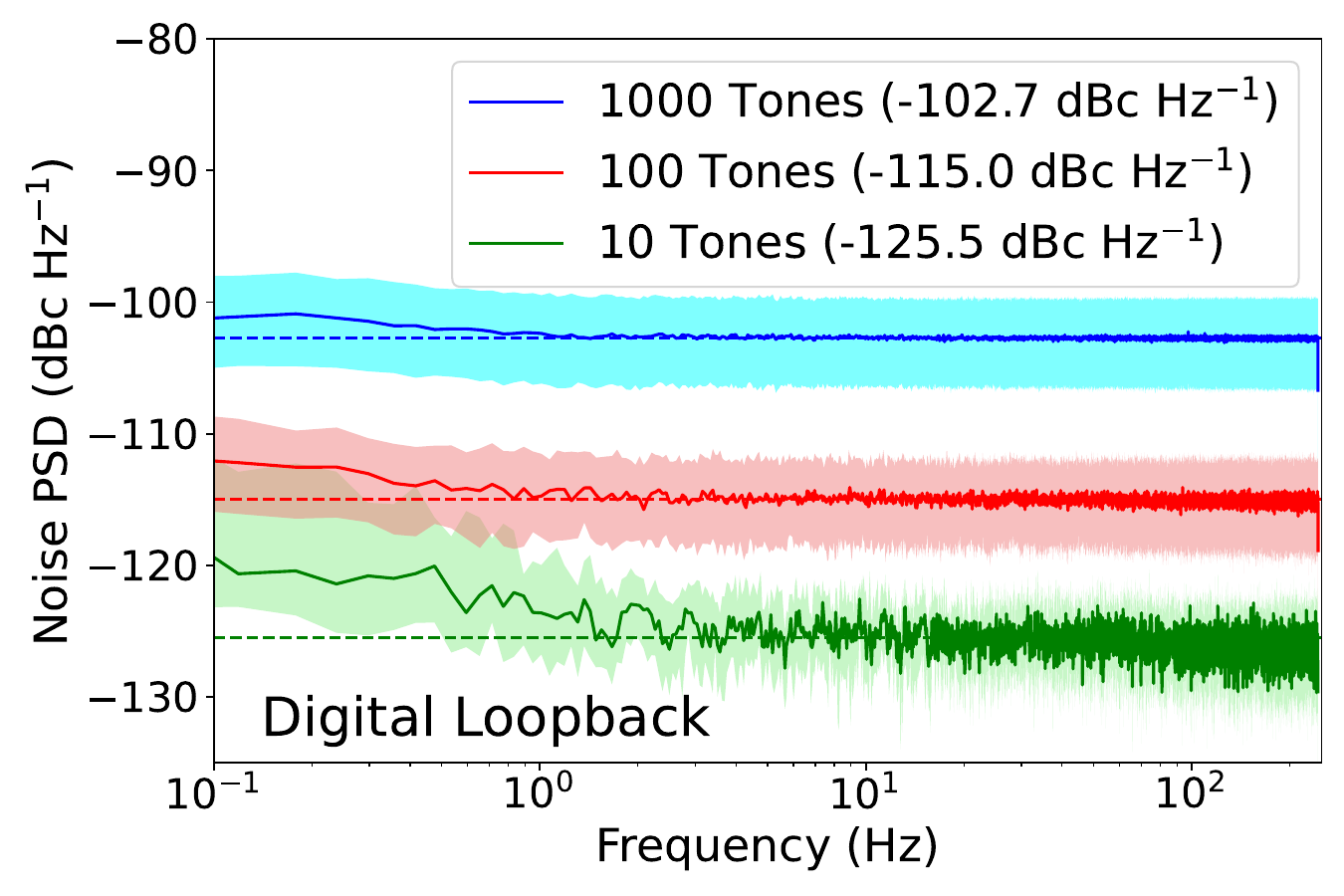}
    \includegraphics[width=0.49\textwidth]{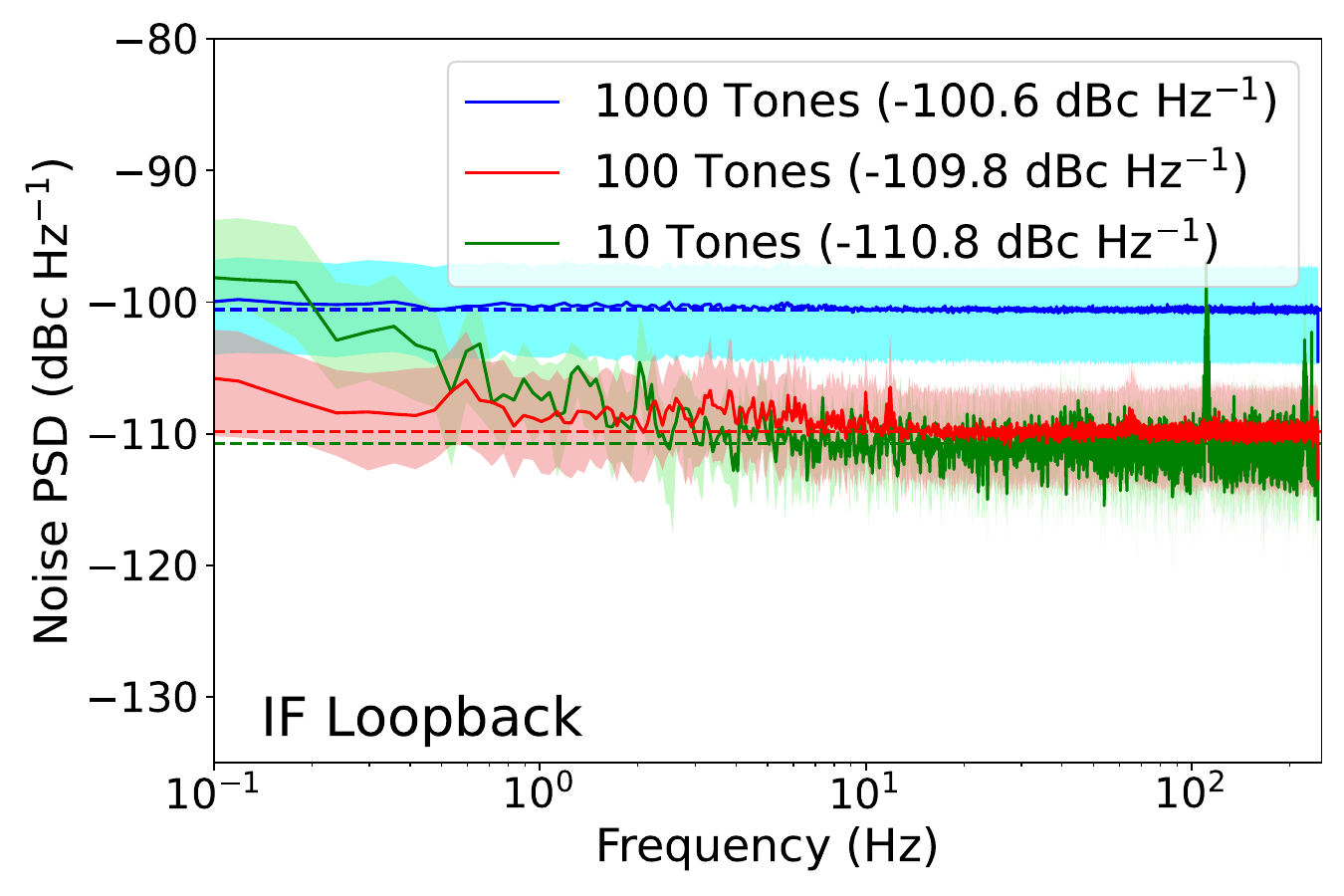}
    \caption{Noise performance of the readout system in digital loopback (left) and IF loopback (right) with 1000 (blue), 100 (red), and 10 (green) tones. Solid lines denote the median tone performance, and shaded areas enclose 68 percent of the tones. The median flat-spectrum noise level is denoted by a dashed line, with the numerical value provided in the caption. In all cases, correlated noise has been removed. Note that correlated noise removal is degraded when only 10 tones are available, resulting in a more steeply rising spectrum at low frequencies and greater residual contamination in the narrow, higher-frequency spectral features compared to cases with 100 or 1000 tones.}
    \label{fig:n_tone_psd}
\end{figure}

Because the noise performance in digital loopback is expected to be limited by the digitization noise of the ADCs (see Figure~\ref{fig:ADC_SNR_SFDR}), and because the DACs are able to generate a fixed total power, the per-tone noise PSD should decrease with the total number of tones since each individual tone will have more power. To test this, we also collected noise data following the same procedures given above while operating with 100 and 10 total tones, with the results shown in Figure~\ref{fig:n_tone_psd}. These data closely match expectations, with a decrease of approximately 10~dBc~Hz$^{-1}$ for each factor of 10 reduction in tone count. Quantitatively, the average noise PSD is $-102.7$~dBc~Hz$^{-1}$, $-115.0$~dBc~Hz$^{-1}$, and $-125.5$~dBc~Hz$^{-1}$ with 1000, 100, and 10 tones. We note that there is a slight degradation in the removal of low-frequency noise with fewer tones, resulting in a more strongly rising spectrum relative to the flat-spectrum noise level. Also, by directly connecting the DAC output to the ADC input, we expect the noise of the total system to be approximately $-147$~dBm~Hz$^{-1}$ based on the independent ADC and DAC noise measurements described in Section~\ref{sec:DAC_noise}. Given the typical per-tone power of $-44$~dBm when outputting 1000 tones, we thus expect a noise level of $-103$~dBc~Hz$^{-1}$, consistent with our average measured value of $-102.7$~dBc~Hz$^{-1}$.

\subsection{IF Loopback}
\label{sec:IF_noise}

With the noise performance of the digital electronics established, we next add the IF system and connect its output directly to its input. We refer to operation in this state as IF loopback. We again utilize the same tone lists as described above for digital loopback, and our baseline is to set the LO to 400~MHz.

First, considering the raw data shown in Figure~\ref{fig:raw_psd}, we note that the primary change compared to digital loopback is the addition of noise in relatively narrow spectral features at high frequencies above approximately 100~Hz. As with the rising noise PSD at low frequency due to the digital electronics, these narrow spectral features are highly correlated among all the tones. By forming templates from the 8--10 highest amplitude principal components, corresponding to 6--8 new templates compared to the digital loopback analysis, it is again possible to obtain noise with an approximately flat spectrum. The amplitude of this spectrum is $-100.6$~dBc~Hz$^{-1}$, indicating that the IF system adds approximately 2~dB of noise relative to the digital system.

We further assess the performance in the IF loopback for 100 and 10 tones. When going from 1000 tones to 100 tones, the noise improves by approximately 10~dB to $-109.8$~dBc~Hz$^{-1}$, although there is little change when further decreasing to 10 tones. This suggests that the IF system introduces a noise floor near $-110$~dBc~Hz$^{-1}$. We attribute this noise floor to imperfect cancellation of LO phase noise in the homodyne mixing setup of the IF system, which uses the same LO for both upconversion and downconversion of the tones between the digital system and the IF system's input and output. At the LO frequency of 400~MHz, and within the relevant bandwidth of approximately 100~Hz relative to this LO frequency, the phase noise of the LO module utilized in the IF system is $-85$~dBc~Hz$^{-1}$, suggesting that the homodyne mixing reduces this noise by approximately 25~dB.

\subsection{KIDs}

\begin{figure}
    \centering
    \includegraphics[width=0.49\textwidth]{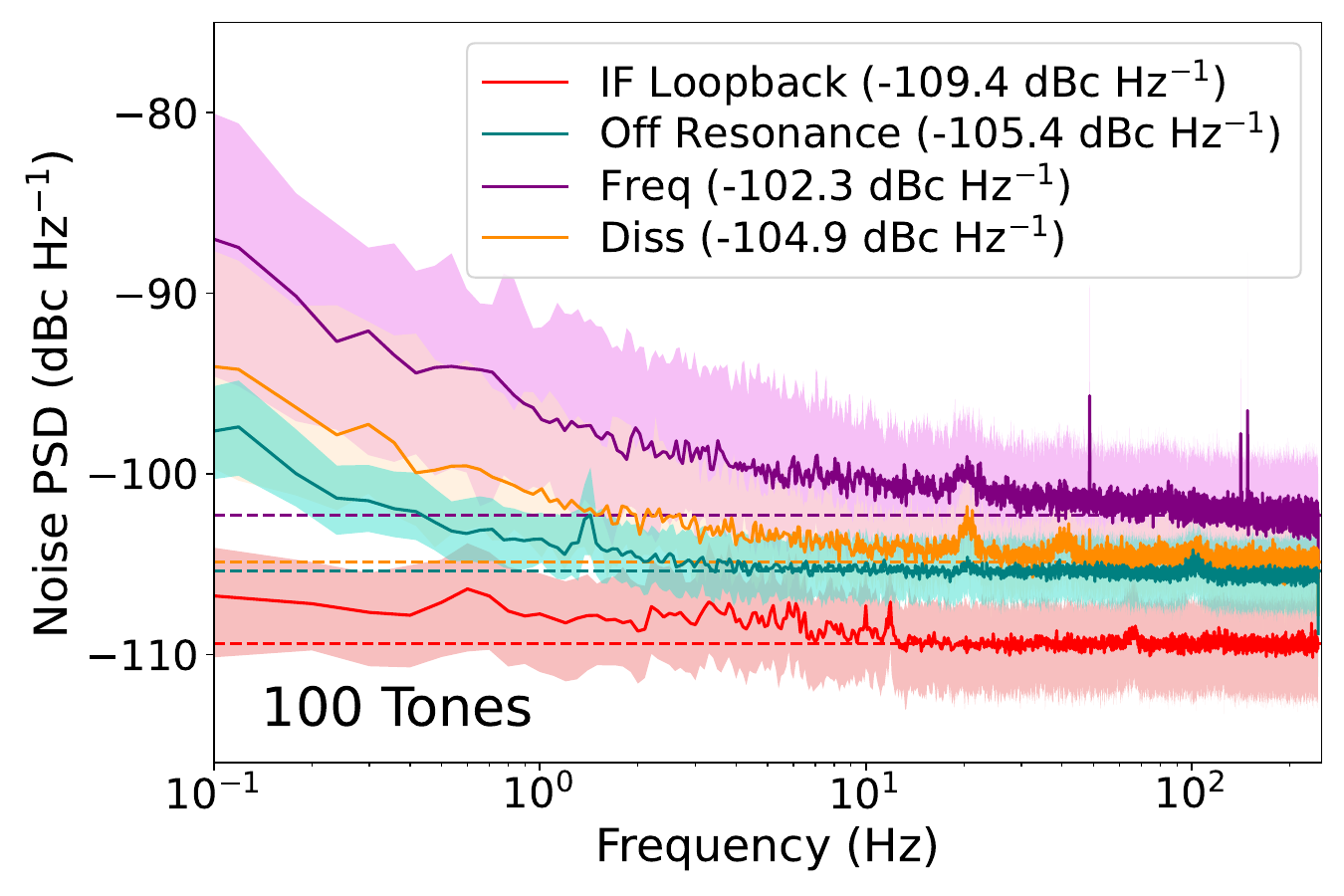}
    \includegraphics[width=0.49\textwidth]{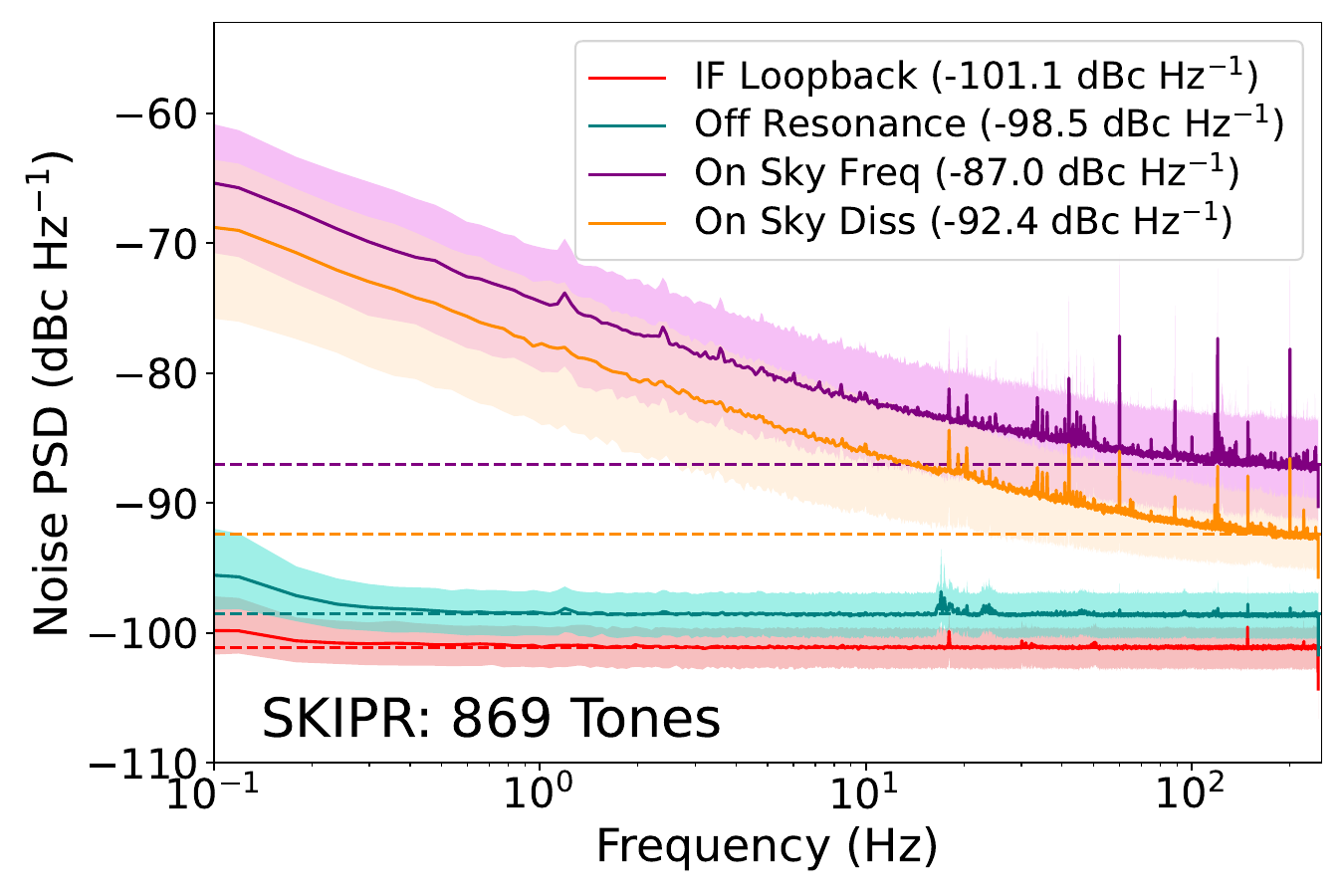}
    \caption{Noise performance of the readout system for two typical use cases. Shown on the left is a 100-tone measurement of a prototype KID array setup in a dark configuration, while the right plot corresponds to an on-sky KID detector array in the SKIPR instrument with 869 readout tones. In both plots, noise is shown for IF loopback (red), loopback through the cryogenic system but bypassing the KIDs (aqua), and with tones centered on the KID resonators and referenced to the frequency and dissipation bases (purple and orange). For both examples, the noise contributed by the readout system is well below the detector noise.}
    \label{fig:telescope_psd}
\end{figure}

To better establish the performance of the readout system, we also consider two common use cases. The first relates to characterizing a prototype KID array with of order 100 detectors in a lab-based configuration fully enclosed within a cryogenic chamber (i.e., a ``dark'' setup). The noise performance of these detectors, which are intended for ground-based mm-wave and submm photometry, has already been characterized by \citet{Hempel-Costello2025}. Thus, they provide a well-established test case. We first determine the noise performance of the readout system through the full cryogenic system ``off resonance'', i.e., bypassing the KIDs, finding a flat-spectrum noise amplitude of $-105.4$~dBc~Hz$^{-1}$, see Figure~\ref{fig:telescope_psd}. This is approximately $4$~dB higher than the value found from the 100-tone IF loopback test. Among the components in the cryogenic system, we expect the LNA to contribute the most noise, so we estimate its amplitude. Given the typical power level at the input to this LNA ($\gtrsim -80$~dBm), and its measured noise temperature ($\simeq 4.5$~K), it should contribute $\lesssim -110$~dBc~Hz$^{-1}$. This is nearly identical to the measured noise of the readout system in IF loopback, suggesting that the combined system should have a noise amplitude of approximately $-107$~dBc~Hz$^{-1}$. While this is slightly lower than the measured off-resonance noise amplitude, the 1--2~dB difference is comparable to, or better than, the precision of our estimated power level at the LNA input. Thus, we conclude that the measured performance of the cryogenic testbed is more likely limited by the LNA rather than the room-temperature readout system.

We also measure noise with readout tones centered on KID resonances, finding typical flat-spectrum noise amplitudes of $-102.3$~dBc~Hz$^{-1}$ and $-104.9$~dBc~Hz$^{-1}$ in the frequency and dissipation detector response. The noise associated with the dissipation response is nearly identical to the value measured off-resonance, indicating that it is limited by the cryogenic LNA. In contrast, the frequency response, which is far more sensitive to the desired astronomical signals, has noise performance limited by the detector. We note that, for this test, only 100 tones were utilized. If we instead used 1000 readout tones, the room-temperature readout would have a flat-spectrum noise amplitude approximately 1--2~dB higher than the detector noise for the KID frequency response. While this could affect lab-based noise characterization of large arrays of KIDs in a dark configuration, the expected KID noise on-sky due to the random arrival of photons is calculated to be approximately 10~dB higher than in the dark lab-based setup \citep{Hempel-Costello2025}. Therefore, the readout system's performance is sufficient for large arrays of such detectors in normal on-sky operation.

For our second test case, we consider a SKIPR KID array with 869 detectors operating in an on-sky configuration \citep{Sayers2025}, see Figure~\ref{fig:telescope_psd}. We again quantify the noise performance off-resonance, where we find a flat-spectrum noise amplitude of $-98.5$~dBc~Hz$^{-1}$, which is approximately 2--3~dB higher than the IF loopback noise amplitude for a similar number of tones. As noted in Sec.~\ref{sec:requirements}, the cryogenic LNA in the SKIPR system is expected to contribute a noise amplitude of approximately $-100$~dBc~Hz$^{-1}$. Combined with the measured IF loopback noise of $-100.6$~dBc~Hz$^{-1}$, the expected off-resonance noise is approximately $-97$~dBc~Hz$^{-1}$, consistent within 1--2~dB of the measured noise value. When the tones are centered on the KID resonators, the flat-spectrum noise amplitude increases to approximately $-87$~dBc~Hz$^{-1}$ in the frequency response direction and approximately $-92$~dBc~Hz$^{-1}$ in the dissipation response direction, suggesting that the overall performance is detector-noise limited.

\section{Summary}

Based on the Xilinx ZCU111 RFSoC evaluation board, we have demonstrated a readout system capable of probing up to 2048 resonances per board. Two separate channels, each with 512~MHz of bandwidth, can independently readout 1024 resonators. We combine this digital readout system with a custom IF system capable of converting the digital band up to frequencies as high as 4 GHz. To facilitate interaction with both the digital and IF systems, we have also developed a software interface that can set readout parameters and perform a range of analyses typical of KIDs used in millimeter-wave imagers. The measured noise performance of the individual components is consistent with expectations based on their datasheets, and the full-system noise performance is sufficiently low to have a negligible impact on typical KID-based imagers operating in the millimeter-wavelength regime. In sum, this package provides a readily scalable solution to readout the large KID arrays in existing and planned imagers. It is also applicable to a wide range of other applications involving high-quality-factor resonators.

\section*{Acknowledgments}

This work was supported by the Office of Naval Research Award Number N000142512109.

\bibliographystyle{ws-jai}
\bibliography{sample}

@ARTICLE{Adam2018,
       author = {{Adam}, R. and {Adane}, A. and {Ade}, P.~A.~R. and {Andr{\'e}}, P. and {Andrianasolo}, A. and {Aussel}, H. and {Beelen}, A. and {Beno{\^\i}t}, A. and {Bideaud}, A. and {Billot}, N. and {Bourrion}, O. and {Bracco}, A. and {Calvo}, M. and {Catalano}, A. and {Coiffard}, G. and {Comis}, B. and {De Petris}, M. and {D{\'e}sert}, F.-X. and {Doyle}, S. and {Driessen}, E.~F.~C. and {Evans}, R. and {Goupy}, J. and {Kramer}, C. and {Lagache}, G. and {Leclercq}, S. and {Leggeri}, J.-P. and {Lestrade}, J.-F. and {Mac{\'\i}as-P{\'e}rez}, J.~F. and {Mauskopf}, P. and {Mayet}, F. and {Maury}, A. and {Monfardini}, A. and {Navarro}, S. and {Pascale}, E. and {Perotto}, L. and {Pisano}, G. and {Ponthieu}, N. and {Rev{\'e}ret}, V. and {Rigby}, A. and {Ritacco}, A. and {Romero}, C. and {Roussel}, H. and {Ruppin}, F. and {Schuster}, K. and {Sievers}, A. and {Triqueneaux}, S. and {Tucker}, C. and {Zylka}, R.},
        title = "{The NIKA2 large-field-of-view millimetre continuum camera for the 30 m IRAM telescope}",
      journal = {\aap},
         year = 2018,
        month = jan,
       volume = {609},
          eid = {A115},
        pages = {A115},
          doi = {10.1051/0004-6361/201731503},
archivePrefix = {arXiv},
       eprint = {1707.00908},
 primaryClass = {astro-ph.IM},
       adsurl = {https://ui.adsabs.harvard.edu/abs/2018A&A...609A.115A}
}

@INPROCEEDINGS{Baudry2012,
       author = {{Baudry}, Alain and {Lacasse}, Richard and {Escoffier}, Ray and {Webber}, John and {Greenberg}, Joseph and {Platt}, Laurence and {Treacy}, Robert and {Saez}, Alejandro F. and {Cais}, Philippe and {Comoretto}, Giovanni and {Quertier}, Benjamin and {Okumura}, Sachiko K. and {Kamazaki}, Takeshi and {Chikada}, Yoshihiro and {Watanabe}, Manabu and {Okuda}, Takeshi and {Kurono}, Yasutake and {Iguchi}, Satoru},
        title = "{Performance highlights of the ALMA correlators}",
    booktitle = {Millimeter, Submillimeter, and Far-Infrared Detectors and Instrumentation for Astronomy VI},
         year = 2012,
       editor = {{Holland}, Wayne S. and {Zmuidzinas}, Jonas},
       series = {Society of Photo-Optical Instrumentation Engineers (SPIE) Conference Series},
       volume = {8452},
        month = sep,
          eid = {845217},
        pages = {845217},
          doi = {10.1117/12.925700},
       adsurl = {https://ui.adsabs.harvard.edu/abs/2012SPIE.8452E..17B}
}

@ARTICLE{Bourrion2016,
       author = {{Bourrion}, O. and {Benoit}, A. and {Bouly}, J.~L. and {Bouvier}, J. and {Bosson}, G. and {Calvo}, M. and {Catalano}, A. and {Goupy}, J. and {Li}, C. and {Mac{\'\i}as-P{\'e}rez}, J.~F. and {Monfardini}, A. and {Tourres}, D. and {Ponchant}, N. and {Vescovi}, C.},
        title = "{NIKEL\_AMC: Readout electronics for the NIKA2 experiment}",
      journal = {arXiv e-prints},
         year = 2016,
        month = feb,
          eid = {arXiv:1602.01288},
        pages = {arXiv:1602.01288},
          doi = {10.48550/arXiv.1602.01288},
archivePrefix = {arXiv},
       eprint = {1602.01288},
 primaryClass = {astro-ph.IM},
       adsurl = {https://ui.adsabs.harvard.edu/abs/2016arXiv160201288B}
}

@INPROCEEDINGS{Duan2010,
       author = {{Duan}, Ran and {McHugh}, Sean and {Serfass}, Bruno and {Mazin}, Benjamin A. and {Merrill}, A. and {Golwala}, Sunil R. and {Downes}, Thomas P. and {Czakon}, Nicole G. and {Day}, Peter K. and {Gao}, Jiansong and {Glenn}, Jason and {Hollister}, Matthew I. and {Leduc}, Henry G. and {Maloney}, Philip R. and {Noroozian}, Omid and {Nguyen}, Hien T. and {Sayers}, Jack and {Schlaerth}, James A. and {Siegel}, Seth and {Vaillancourt}, John E. and {Vayonakis}, Anastasios and {Wilson}, Philip R. and {Zmuidzinas}, Jonas},
        title = "{An open-source readout for MKIDs}",
    booktitle = {Millimeter, Submillimeter, and Far-Infrared Detectors and Instrumentation for Astronomy V},
         year = 2010,
       editor = {{Holland}, Wayne S. and {Zmuidzinas}, Jonas},
       series = {Society of Photo-Optical Instrumentation Engineers (SPIE) Conference Series},
       volume = {7741},
        month = jul,
          eid = {77411V},
        pages = {77411V},
          doi = {10.1117/12.856832},
       adsurl = {https://ui.adsabs.harvard.edu/abs/2010SPIE.7741E..1VD}
}

@ARTICLE{EBEX2018,
       author = {{EBEX Collaboration} and {Abitbol}, Maximilian and {Aboobaker}, Asad M. and {Ade}, Peter and {Araujo}, Derek and {Aubin}, Fran{\c{c}}ois and {Baccigalupi}, Carlo and {Bao}, Chaoyun and {Chapman}, Daniel and {Didier}, Joy and {Dobbs}, Matt and {Feeney}, Stephen M. and {Geach}, Christopher and {Grainger}, Will and {Hanany}, Shaul and {Helson}, Kyle and {Hillbrand}, Seth and {Hilton}, Gene and {Hubmayr}, Johannes and {Irwin}, Kent and {Jaffe}, Andrew and {Johnson}, Bradley and {Jones}, Terry and {Klein}, Jeff and {Korotkov}, Andrei and {Lee}, Adrian and {Levinson}, Lorne and {Limon}, Michele and {MacDermid}, Kevin and {Miller}, Amber D. and {Milligan}, Michael and {Raach}, Kate and {Reichborn-Kjennerud}, Britt and {Reintsema}, Carl and {Sagiv}, Ilan and {Smecher}, Graeme and {Tucker}, Gregory S. and {Westbrook}, Benjamin and {Young}, Karl and {Zilic}, Kyle},
        title = "{The EBEX Balloon-borne Experiment{\textemdash}Detectors and Readout}",
      journal = {\apjs},
         year = 2018,
        month = nov,
       volume = {239},
       number = {1},
          eid = {8},
        pages = {8},
          doi = {10.3847/1538-4365/aae436},
archivePrefix = {arXiv},
       eprint = {1803.01018},
 primaryClass = {astro-ph.IM},
       adsurl = {https://ui.adsabs.harvard.edu/abs/2018ApJS..239....8E}
}

@INPROCEEDINGS{Filippini2010,
       author = {{Filippini}, J.~P. and {Ade}, P.~A.~R. and {Amiri}, M. and {Benton}, S.~J. and {Bihary}, R. and {Bock}, J.~J. and {Bond}, J.~R. and {Bonetti}, J.~A. and {Bryan}, S.~A. and {Burger}, B. and {Chiang}, H.~C. and {Contaldi}, C.~R. and {Crill}, B.~P. and {Dor{\'e}}, O. and {Farhang}, M. and {Fissel}, L.~M. and {Gandilo}, N.~N. and {Golwala}, S.~R. and {Gudmundsson}, J.~E. and {Halpern}, M. and {Hasselfield}, M. and {Hilton}, G. and {Holmes}, W. and {Hristov}, V.~V. and {Irwin}, K.~D. and {Jones}, W.~C. and {Kuo}, C.~L. and {MacTavish}, C.~J. and {Mason}, P.~V. and {Montroy}, T.~E. and {Morford}, T.~A. and {Netterfield}, C.~B. and {O'Dea}, D.~T. and {Rahlin}, A.~S. and {Reintsema}, C.~D. and {Ruhl}, J.~E. and {Runyan}, M.~C. and {Schenker}, M.~A. and {Shariff}, J.~A. and {Soler}, J.~D. and {Trangsrud}, A. and {Tucker}, C. and {Tucker}, R.~S. and {Turner}, A.~D.},
        title = "{SPIDER: a balloon-borne CMB polarimeter for large angular scales}",
    booktitle = {Millimeter, Submillimeter, and Far-Infrared Detectors and Instrumentation for Astronomy V},
         year = 2010,
       editor = {{Holland}, Wayne S. and {Zmuidzinas}, Jonas},
       series = {Society of Photo-Optical Instrumentation Engineers (SPIE) Conference Series},
       volume = {7741},
        month = jul,
          eid = {77411N},
        pages = {77411N},
          doi = {10.1117/12.857720},
archivePrefix = {arXiv},
       eprint = {1106.2158},
 primaryClass = {astro-ph.CO},
       adsurl = {https://ui.adsabs.harvard.edu/abs/2010SPIE.7741E..1NF}
}

@ARTICLE{Gordon2016,
       author = {{Gordon}, Samuel and {Dober}, Brad and {Sinclair}, Adrian and {Rowe}, Samuel and {Bryan}, Sean and {Mauskopf}, Philip and {Austermann}, Jason and {Devlin}, Mark and {Dicker}, Simon and {Gao}, Jiansong and {Hilton}, Gene C. and {Hubmayr}, Johannes and {Jones}, Glenn and {Klein}, Jeffrey and {Lourie}, Nathan P. and {McKenney}, Christopher and {Nati}, Federico and {Soler}, Juan D. and {Strader}, Matthew and {Vissers}, Michael},
        title = "{An Open Source, FPGA-Based LeKID Readout for BLAST-TNG: Pre-Flight Results}",
      journal = {Journal of Astronomical Instrumentation},
         year = 2016,
        month = dec,
       volume = {5},
       number = {4},
          eid = {1641003},
        pages = {1641003},
          doi = {10.1142/S2251171716410038},
archivePrefix = {arXiv},
       eprint = {1611.05400},
 primaryClass = {astro-ph.IM},
       adsurl = {https://ui.adsabs.harvard.edu/abs/2016JAI.....541003G}
}

@ARTICLE{Hattori2016,
       author = {{Hattori}, K. and {Akiba}, Y. and {Arnold}, K. and {Barron}, D. and {Bender}, A.~N. and {Cukierman}, A. and {de Haan}, T. and {Dobbs}, M. and {Elleflot}, T. and {Hasegawa}, M. and {Hazumi}, M. and {Holzapfel}, W. and {Hori}, Y. and {Keating}, B. and {Kusaka}, A. and {Lee}, A. and {Montgomery}, J. and {Rotermund}, K. and {Shirley}, I. and {Suzuki}, A. and {Whitehorn}, N.},
        title = "{Development of Readout Electronics for POLARBEAR-2 Cosmic Microwave Background Experiment}",
      journal = {Journal of Low Temperature Physics},
         year = 2016,
        month = jul,
       volume = {184},
       number = {1-2},
        pages = {512-518},
          doi = {10.1007/s10909-015-1448-x},
archivePrefix = {arXiv},
       eprint = {1512.07663},
 primaryClass = {astro-ph.IM},
       adsurl = {https://ui.adsabs.harvard.edu/abs/2016JLTP..184..512H}
}

@ARTICLE{Heaton2023,
       author = {{Heaton}, Grigory and {Cook}, Walter and {Bock}, James and {Burnham}, Jill and {Condon}, Sam and {Hristov}, Viktor and {Hui}, Howard and {Kecman}, Branislav and {Korngut}, Phillip and {Miyasaka}, Hiromasa and {Nguyen}, Chi and {Padin}, Stephen and {Viero}, Marco},
        title = "{Noise Reduction Methods for Large-scale Intensity-mapping Measurements with Infrared Detector Arrays}",
      journal = {\apjs},
         year = 2023,
        month = oct,
       volume = {268},
       number = {2},
          eid = {44},
        pages = {44},
          doi = {10.3847/1538-4365/acebc1},
archivePrefix = {arXiv},
       eprint = {2309.15966},
 primaryClass = {astro-ph.IM},
       adsurl = {https://ui.adsabs.harvard.edu/abs/2023ApJS..268...44H}
}

@ARTICLE{Hempel-Costello2025,
       author = {{Hempel-Costello}, Simon and {Beyer}, Andrew D. and {Cunnane}, Dan and {Day}, Peter K. and {Defrance}, Fabien and {Frez}, Cliff and {Gavidia}, Adriana and {Golwala}, Sunil R. and {Kim}, Junhan and {Martin}, Jean-Marc and {Sadou}, Yann and {Sayers}, Jack and {Shu}, Shibo and {Yu}, Shiling},
        title = "{Low-Frequency Noise Performance of Microstrip-Coupled Lumped-Element Aluminum KIDs using Hydrogenated Amorphous Silicon Parallel-Plate Capacitors for NEW-MUSIC}",
      journal = {arXiv e-prints},
         year = 2025,
        month = nov,
          eid = {arXiv:2511.08898},
        pages = {arXiv:2511.08898},
          doi = {10.48550/arXiv.2511.08898},
archivePrefix = {arXiv},
       eprint = {2511.08898},
 primaryClass = {astro-ph.IM},
       adsurl = {https://ui.adsabs.harvard.edu/abs/2025arXiv251108898H}
}

@BOOK{Jolliffe1986,
       author = {{Jolliffe}, I.~T.},
        title = "{Principal component analysis}",
         year = 1986,
         publisher = {springer},
       adsurl = {https://ui.adsabs.harvard.edu/abs/1986pca..book.....J}
}

@INPROCEEDINGS{Paiella2019,
       author = {{Paiella}, A. and {Battistelli}, E.~S. and {Castellano}, M.~G. and {Colantoni}, I. and {Columbro}, F. and {Coppolecchia}, A. and {D'Alessandro}, G. and {de Bernardis}, P. and {Gordon}, S. and {Lamagna}, L. and {Mani}, H. and {Masi}, S. and {Mauskopf}, P. and {Pettinari}, G. and {Piacentini}, F. and {Presta}, G.},
        title = "{Kinetic Inductance Detectors and readout electronics for the OLIMPO experiment}",
    booktitle = {Journal of Physics Conference Series},
         year = 2019,
       series = {Journal of Physics Conference Series},
       volume = {1182},
        month = feb,
    publisher = {IOP},
          eid = {012005},
        pages = {012005},
          doi = {10.1088/1742-6596/1182/1/012005},
archivePrefix = {arXiv},
       eprint = {1904.01890},
 primaryClass = {astro-ph.IM},
       adsurl = {https://ui.adsabs.harvard.edu/abs/2019JPhCS1182a2005P}
}

@ARTICLE{Reyes2026,
       author = {{Reyes}, N. and {Weiss}, A. and {Yates}, S.~J.~C. and {Baryshev}, A.~M. and {C{\'a}mara-Mayorga}, I. and {Dabironezare}, S. and {Endo}, A. and {Ferrari}, L. and {G{\"o}rlitz}, A. and {Grutzeck}, G. and {G{\"u}sten}, R. and {Heiter}, C. and {Heyminck}, S. and {Hochg{\"u}rtel}, S. and {Hoevers}, H. and {Jorquera}, S. and {Kov{\'a}cs}, A. and {Koopmans}, D. and {K{\"o}nig}, C. and {Llombart}, N. and {Menten}, K.~M. and {Murugesan}, V. and {Ridder}, M. and {Schmitz}, A. and {Thoen}, D.~J. and {van der Linden}, A.~J. and {Wang}, L. and {Yurduseven}, O. and {Baselmans}, J.~J.~A. and {Klein}, B.},
        title = "{AMKID: A large KID-based camera at the APEX telescope}",
      journal = {\aap},
         year = 2026,
        month = mar,
       volume = {707},
          eid = {A294},
        pages = {A294},
          doi = {10.1051/0004-6361/202558596},
archivePrefix = {arXiv},
       eprint = {2512.14905},
 primaryClass = {astro-ph.IM},
       adsurl = {https://ui.adsabs.harvard.edu/abs/2026A&A...707A.294R}
}

@ARTICLE{Rowe2023,
       author = {{Rowe}, S. and {Tapia}, M. and {Barry}, P.~S. and {Karkare}, K.~S. and {Papageorgiou}, A. and {Ade}, P.~A.~R. and {Brien}, T.~L.~R. and {Castillo-Dom{\'\i}nguez}, E. and {Ferrusca}, D. and {G{\'o}mez-Rivera}, V. and {Hargrave}, P. and {Hern{\'a}ndez-Rebollar}, J.~L. and {Hornsby}, A. and {J{\'a}uregui-Garc{\'\i}a}, J.~M. and {Mauskopf}, P. and {Murias}, D. and {Pascale}, E. and {P{\'e}rez}, A. and {Smith}, M.~W.~L. and {Tucker}, C. and {Vel{\'a}zquez}, M. and {Ventura}, S. and {Hughes}, D.~H. and {Doyle}, S.},
        title = "{The MUSCAT Readout Electronics Backend: Design and Pre-deployment Performance}",
      journal = {Journal of Low Temperature Physics},
         year = 2023,
        month = jun,
       volume = {211},
       number = {5-6},
        pages = {289-301},
          doi = {10.1007/s10909-022-02868-9},
       adsurl = {https://ui.adsabs.harvard.edu/abs/2023JLTP..211..289R}
}

@ARTICLE{Sayers2025,
       author = {{Sayers}, J. and {Cunnane}, D. and {Crystian}, S. and {Day}, P.~K. and {Defrance}, F. and {Eom}, B.~H. and {Greenfield}, J. and {Hollister}, M. and {Johnson}, B.~R. and {LeDuc}, H.~G. and {Mauskopf}, P. and {McNichols}, N. and {Roberson}, C. and {Runyan}, M.~C. and {Sriram}, A.~B. and {Stanton}, S. and {Stephenson}, R.~C. and {Walters}, L.~C. and {Weeks}, E.},
        title = "{A millimeter-wave photometric camera for long-range imaging through optical obscurants using kinetic inductance detectors}",
      journal = {Review of Scientific Instruments},
         year = 2025,
        month = mar,
       volume = {96},
       number = {3},
          eid = {034501},
        pages = {034501},
          doi = {10.1063/5.0249704},
archivePrefix = {arXiv},
       eprint = {2502.14607},
 primaryClass = {astro-ph.IM},
       adsurl = {https://ui.adsabs.harvard.edu/abs/2025RScI...96c4501S}
}

@ARTICLE{Schillaci2023,
       author = {{Schillaci}, Alessandro and {Ade}, P.~A.~R. and {Ahmed}, Z. and {Amiri}, M. and {Barkats}, D. and {Basu Thakur}, R. and {Bischoff}, C.~A. and {Beck}, D. and {Bock}, J.~J. and {Buza}, V. and {Cheshire}, J. and {Connors}, J. and {Cornelison}, J. and {Crumrine}, M. and {Cukierman}, A. and {Denison}, E. and {Dierickx}, M. and {Duband}, L. and {Eiben}, M. and {Fatigoni}, S. and {Filippini}, J.~P. and {Giannakopoulos}, C. and {Goeckner-Wald}, N. and {Goldfinger}, D. and {Grayson}, J.~A. and {Grimes}, P. and {Hall}, G. and {Halal}, G. and {Halpern}, M. and {Hand}, E. and {Harrison}, S. and {Henderson}, S. and {Hildebrandt}, S.~R. and {Hilton}, G.~C. and {Hubmayr}, J. and {Hui}, H. and {Irwin}, K.~D. and {Kang}, J. and {Karkare}, K.~S. and {Kefeli}, S. and {Kovac}, J.~M. and {Kuo}, C.~L. and {Lau}, K. and {Leitch}, E.~M. and {Lennox}, A. and {Megerian}, K.~G. and {Miller}, O.~Y. and {Minutolo}, L. and {Moncelsi}, L. and {Nakato}, Y. and {Namikawa}, T. and {Nguyen}, H.~T. and {O'Brient}, R. and {Palladino}, S. and {Petroff}, M. and {Precup}, N. and {Prouve}, T. and {Pryke}, C. and {Racine}, B. and {Reintsema}, C.~D. and {Schmitt}, B.~L. and {Singari}, B. and {Soliman}, A. and {Germaine}, T. St. and {Steinbach}, B. and {Sudiwala}, R.~V. and {Thompson}, K.~L. and {Tucker}, C. and {Turner}, A.~D. and {Umilt{\`a}}, C. and {Verges}, C. and {Vieregg}, A.~G. and {Wandui}, A. and {Weber}, A.~C. and {Wiebe}, D.~V. and {Willmert}, J. and {Wu}, W.~L.~K. and {Yang}, E. and {Yoon}, K.~W. and {Young}, E. and {Yu}, C. and {Zeng}, L. and {Zhang}, C. and {Zhang}, S.},
        title = "{BICEP Array: 150 GHz Detector Module Development}",
      journal = {Journal of Low Temperature Physics},
         year = 2023,
        month = dec,
       volume = {213},
       number = {5-6},
        pages = {317-326},
          doi = {10.1007/s10909-023-03005-w},
archivePrefix = {arXiv},
       eprint = {2111.14785},
 primaryClass = {astro-ph.IM},
       adsurl = {https://ui.adsabs.harvard.edu/abs/2023JLTP..213..317S}
}

@INPROCEEDINGS{Sinclair2022,
       author = {{Sinclair}, Adrian K. and {Stephenson}, Ryan C. and {Roberson}, Cody A. and {Weeks}, Eric L. and {Burgoyne}, James and {Huber}, Anthony I. and {Mauskopf}, Philip M. and {Chapman}, Scott C. and {Austermann}, Jason E. and {Choi}, Steve K. and {Duell}, Cody J. and {Fich}, Michel and {Groppi}, Christopher E. and {Huber}, Zachary and {Niemack}, Michael D. and {Nikola}, Thomas and {Rossi}, Kayla M. and {Sriram}, Adhitya and {Stacey}, Gordon J. and {Szakiel}, Erik and {Tsuchitori}, Joel and {Vavagiakis}, Eve M. and {Wheeler}, Jordan D.},
        title = "{CCAT-prime: RFSoC based readout for frequency multiplexed kinetic inductance detectors}",
    booktitle = {Millimeter, Submillimeter, and Far-Infrared Detectors and Instrumentation for Astronomy XI},
         year = 2022,
       editor = {{Zmuidzinas}, Jonas and {Gao}, Jian-Rong},
       series = {Society of Photo-Optical Instrumentation Engineers (SPIE) Conference Series},
       volume = {12190},
        month = aug,
          eid = {121900W},
        pages = {121900W},
          doi = {10.1117/12.2629722},
archivePrefix = {arXiv},
       eprint = {2208.07465},
 primaryClass = {astro-ph.IM},
       adsurl = {https://ui.adsabs.harvard.edu/abs/2022SPIE12190E..0WS}
}

@INPROCEEDINGS{Sinclair2024,
       author = {{Sinclair}, Adrian K. and {Burgoyne}, James and {Huber}, Anthony I. and {Murphy}, Colin and {Choi}, Steve K. and {Duell}, Cody J. and {Huber}, Zachary B. and {Li}, Yaqiong and {Chapman}, Scott C. and {Niemack}, Michael D. and {Nikola}, Thomas and {Vavagiakis}, Eve M. and {Walker}, Samantha and {Wheeler}, Jordan D. and {Austermann}, Jason and {Lin}, Lawrence and {Xie}, Ruixuan and {Zou}, Bugao and {Mauskopf}, Philip D.},
        title = "{CCAT: detector noise limited performance of the RFSoC-based readout electronics for mm/sub-mm/far-IR KIDs}",
    booktitle = {Millimeter, Submillimeter, and Far-Infrared Detectors and Instrumentation for Astronomy XII},
         year = 2024,
       editor = {{Zmuidzinas}, Jonas and {Gao}, Jian-Rong},
       series = {Society of Photo-Optical Instrumentation Engineers (SPIE) Conference Series},
       volume = {13102},
        month = aug,
          eid = {131022E},
        pages = {131022E},
          doi = {10.1117/12.3020557},
archivePrefix = {arXiv},
       eprint = {2406.14892},
 primaryClass = {astro-ph.IM},
       adsurl = {https://ui.adsabs.harvard.edu/abs/2024SPIE13102E..2ES}
}

@ARTICLE{Smith2024,
       author = {{Smith}, Jennifer Pearl and {Bailey}, John I. and {Cuda}, Aled and {Zobrist}, Nicholas and {Mazin}, Benjamin A.},
        title = "{MKIDGen3: Energy-resolving, single-photon-counting microwave kinetic inductance detector readout on a radio frequency system-on-chip}",
      journal = {Review of Scientific Instruments},
         year = 2024,
        month = nov,
       volume = {95},
       number = {11},
          eid = {114705},
        pages = {114705},
          doi = {10.1063/5.0225768},
archivePrefix = {arXiv},
       eprint = {2406.09764},
 primaryClass = {physics.ins-det},
       adsurl = {https://ui.adsabs.harvard.edu/abs/2024RScI...95k4705S}
}

@article{Stefanazzi2022,
    author = {Stefanazzi, Leandro and Treptow, Kenneth and Wilcer, Neal and Stoughton, Chris and Bradford, Collin and Uemura, Sho and Zorzetti, Silvia and Montella, Salvatore and Cancelo, Gustavo and Sussman, Sara and Houck, Andrew and Saxena, Shefali and Arnaldi, Horacio and Agrawal, Ankur and Zhang, Helin and Ding, Chunyang and Schuster, David I.},
    title = {The QICK (Quantum Instrumentation Control Kit): Readout and control for qubits and detectors},
    journal = {Review of Scientific Instruments},
    volume = {93},
    number = {4},
    pages = {044709},
    year = {2022},
    month = {04},
    issn = {0034-6748},
    doi = {10.1063/5.0076249},
    url = {https://doi.org/10.1063/5.0076249},
    eprint = {https://pubs.aip.org/aip/rsi/article-pdf/doi/10.1063/5.0076249/19817152/044709_1_online.pdf},
}

@INPROCEEDINGS{Wilson2020,
       author = {{Wilson}, Grant W. and {Abi-Saad}, Sophia and {Ade}, Peter and {Aretxaga}, Itziar and {Austermann}, Jason and {Ban}, Yvonne and {Bardin}, Joseph and {Beall}, James and {Berthoud}, Marc and {Bryan}, Sean and {Bussan}, John and {Castillo}, Edgar and {Chavez}, Miguel and {Contente}, Reid and {DeNigris}, N.~S. and {Dober}, Bradley and {Eiben}, Miranda and {Ferrusca}, Daniel and {Fissel}, Laura and {Gao}, Jiansong and {Golec}, Joseph E. and {Golina}, Robert and {Gomez}, Arturo and {Gordon}, Sam and {Gutermuth}, Robert and {Hilton}, Gene and {Hosseini}, Mohsen and {Hubmayr}, Johannes and {Hughes}, David and {Kuczarski}, Stephen and {Lee}, Dennis and {Lunde}, Emily and {Ma}, Zhiyuan and {Mani}, Hamdi and {Mauskopf}, Philip and {McCrackan}, Michael and {McKenney}, Christopher and {McMahon}, Jeffrey and {Novak}, Giles and {Pisano}, Giampaolo and {Pope}, Alexandra and {Ralston}, Amy and {Rodriguez}, Ivan and {S{\'a}nchez-Arg{\"u}elles}, David and {Schloerb}, F. Peter and {Simon}, Sara and {Sinclair}, Adrian and {Souccar}, Kamal and {Torres Campos}, Ana and {Tucker}, Carole and {Ullom}, Joel and {Van Camp}, Eric and {Van Lanen}, Jeff and {Velazquez}, Miguel and {Vissers}, Michael and {Weeks}, Eric and {Yun}, Min S.},
        title = "{The TolTEC camera: an overview of the instrument and in-lab testing results}",
    booktitle = {Millimeter, Submillimeter, and Far-Infrared Detectors and Instrumentation for Astronomy X},
         year = 2020,
       editor = {{Zmuidzinas}, Jonas and {Gao}, Jian-Rong},
       series = {Society of Photo-Optical Instrumentation Engineers (SPIE) Conference Series},
       volume = {11453},
        month = dec,
          eid = {1145302},
        pages = {1145302},
          doi = {10.1117/12.2562331},
       adsurl = {https://ui.adsabs.harvard.edu/abs/2020SPIE11453E..02W}
}

@ARTICLE{Yu2023,
       author = {{Yu}, Cyndia and {Ahmed}, Zeeshan and {Frisch}, Josef C. and {Henderson}, Shawn W. and {Silva-Feaver}, Max and {Arnold}, Kam and {Brown}, David and {Connors}, Jake and {Cukierman}, Ari J. and {D'Ewart}, J. Mitch and {Dober}, Bradley J. and {Dusatko}, John E. and {Haller}, Gunther and {Herbst}, Ryan and {Hilton}, Gene C. and {Hubmayr}, Johannes and {Irwin}, Kent D. and {Kuo}, Chao-Lin and {Mates}, John A.~B. and {Ruckman}, Larry and {Ullom}, Joel and {Vale}, Leila and {Van Winkle}, Daniel D. and {Vasquez}, Jesus and {Young}, Edward},
        title = "{SLAC microresonator RF (SMuRF) electronics: A tone-tracking readout system for superconducting microwave resonator arrays}",
      journal = {Review of Scientific Instruments},
         year = 2023,
        month = jan,
       volume = {94},
       number = {1},
          eid = {014712},
        pages = {014712},
          doi = {10.1063/5.0125084},
archivePrefix = {arXiv},
       eprint = {2208.10523},
 primaryClass = {physics.ins-det},
       adsurl = {https://ui.adsabs.harvard.edu/abs/2023RScI...94a4712Y}
}

@ARTICLE{Zhu2021,
       author = {{Zhu}, Ningfeng and {Bhandarkar}, Tanay and {Coppi}, Gabriele and {Kofman}, Anna M. and {Orlowski-Scherer}, John L. and {Xu}, Zhilei and {Adachi}, Shunsuke and {Ade}, Peter and {Aiola}, Simone and {Austermann}, Jason and {Bazarko}, Andrew O. and {Beall}, James A. and {Bhimani}, Sanah and {Bond}, J. Richard and {Chesmore}, Grace E. and {Choi}, Steve K. and {Connors}, Jake and {Cothard}, Nicholas F. and {Devlin}, Mark and {Dicker}, Simon and {Dober}, Bradley and {Duell}, Cody J. and {Duff}, Shannon M. and {D{\"u}nner}, Rolando and {Fabbian}, Giulio and {Galitzki}, Nicholas and {Gallardo}, Patricio A. and {Golec}, Joseph E. and {Haridas}, Saianeesh K. and {Harrington}, Kathleen and {Healy}, Erin and {Ho}, Shuay-Pwu Patty and {Huber}, Zachary B. and {Hubmayr}, Johannes and {Iuliano}, Jeffrey and {Johnson}, Bradley R. and {Keating}, Brian and {Kiuchi}, Kenji and {Koopman}, Brian J. and {Lashner}, Jack and {Lee}, Adrian T. and {Li}, Yaqiong and {Limon}, Michele and {Link}, Michael and {Lucas}, Tammy J. and {McCarrick}, Heather and {Moore}, Jenna and {Nati}, Federico and {Newburgh}, Laura B. and {Niemack}, Michael D. and {Pierpaoli}, Elena and {Randall}, Michael J. and {Sarmiento}, Karen Perez and {Saunders}, Lauren J. and {Seibert}, Joseph and {Sierra}, Carlos and {Sonka}, Rita and {Spisak}, Jacob and {Sutariya}, Shreya and {Tajima}, Osamu and {Teply}, Grant P. and {Thornton}, Robert J. and {Tsan}, Tran and {Tucker}, Carole and {Ullom}, Joel and {Vavagiakis}, Eve M. and {Vissers}, Michael R. and {Walker}, Samantha and {Westbrook}, Benjamin and {Wollack}, Edward J. and {Zannoni}, Mario},
        title = "{The Simons Observatory Large Aperture Telescope Receiver}",
      journal = {\apjs},
         year = 2021,
        month = sep,
       volume = {256},
       number = {1},
          eid = {23},
        pages = {23},
          doi = {10.3847/1538-4365/ac0db7},
archivePrefix = {arXiv},
       eprint = {2103.02747},
 primaryClass = {astro-ph.IM},
       adsurl = {https://ui.adsabs.harvard.edu/abs/2021ApJS..256...23Z}
}

\end{document}